\documentclass[10pt, prl]{revtex4}
\usepackage{dcolumn}
\usepackage{amsmath}
\usepackage{graphicx}
\usepackage{rotating}
\usepackage{multirow}
\usepackage{booktabs}
\usepackage{color} 

\date{}

\begin{document}

\begin{widetext}
\begin{flushleft}
{\Large
\textbf{Towards a characterization of behavior-disease models}
}
% Insert Author names, affiliations and corresponding author email.
\\
Nicola Perra$^{1,2,\ast}$, 
Duygu Balcan$^{1,3}$, 
Bruno Gon\c calves $^{1,3}$
Alessandro Vespignani $^{1,3,4}$
\\
\bf{1} Center for Complex Networks and Systems Research, School of Informatics and Computing, Indiana University, Bloomington, IN 47408, USA
\\
\bf{2} Linkalab, Center for the Study of Complex Networks, Cagliari 09129, Sardegna, Italy
\\
\bf{3} Pervasive Technology Institute, Indiana University, Bloomington, IN 47406, USA
\\
\bf{4} Institute for Scientific Interchange (ISI), Viale S. Severo 65, 10133, Torino, Italy
\\
$\ast$ E-mail: Corresponding nperra@indiana.edu
\end{flushleft}

\section*{Abstract}
The last decade saw the advent of increasingly realistic epidemic models that leverage on the availability of highly detailed  census and human mobility data. 
Data-driven models aim at a granularity down to the level of households or single individuals. However, relatively little systematic work has been done to provide coupled behavior-disease models able to close the feedback loop between  behavioral changes triggered in the population by an individual's perception of the disease spread and the actual disease spread itself. While models lacking this coupling can be extremely successful in mild epidemics, they obviously will be  of limited use in situations where social disruption or behavioral alterations are induced in the population by knowledge of the disease. Here we propose a characterization of a set of prototypical mechanisms for self-initiated social distancing induced by local and non-local prevalence-based information available to individuals in the population. We characterize the effects of these mechanisms in the framework of a compartmental scheme that enlarges the basic SIR model by considering separate behavioral classes within the population. The transition of individuals in/out of behavioral classes is coupled with the spreading of the disease and provides a rich phase space with multiple epidemic peaks and tipping points. The class of models presented here can be used in the case of data-driven computational approaches to analyze scenarios of social adaptation and behavioral change.   
\end{widetext}

\section*{Introduction}
Understanding human behavior has long been recognized as one of the keys to understanding  epidemic spreading \cite{ferguson07-1,Funk2010}, 
 which has triggered  intense research activity aimed at including social complexity in epidemiological models. Age structure \cite{wallinga06-1},
 human mobility~\cite{Merler2010, balcan09-2, Colizza2007, Colizza2006, Hollingsworth2006, Cooper2006, Ruan2006, ginelli06-1, Hufnagel2004, Grais2004, Grais2003, Flahault1991, Rvachev1985} and 
very detailed data at the individual level~\cite{LonginiAV,episim,ferguson05-1,chao-2010} are now incorporated in most of the realistic models. However, much remains to be done. Models based on social mobility and behavior \cite{balcan09-1,bajardi09-1} have shown to be valuable tools in the quantitative analysis of the unfolding of the recent H1N1 pandemic  \cite{balcan09-1,bajardi09-1}, but it has become clear that societal reactions coupling behavior and disease spreading can have substantial impact on epidemic spreading \cite{gardner09-1, Funk2010} thus defining limitations of most current modeling approaches~\cite{Alex_science}. Societal reactions can be grouped into different classes. First, there are changes  imposed by authorities through the closure of schools, churches, public offices, and bans on public gatherings \cite{cruzpacheco,Hatchett07-1, bootsma07-1, markel07-1}. Second, individuals self-initiate behavioral changes due to the concern induced by the disease~\cite{Fenichel2011, Chen2011, Poletti2011, funk09-1, poletti09-1, Coelho2009, epstein08-1, Bagnoli2007}. Behavioral changes vary from simply  avoiding social contact with infected individuals and crowded spaces~\cite{Crosby1990} to reducing travel~\cite{Campbell1995, Lau2005} and preventing children from attending school. In all cases we have a modification of the spreading process due to the change of mobility or contact patterns in the population.  In general, these behavioral changes may have a considerable impact on  epidemic progression such as the reduction in  epidemic size and  delay of the epidemic peak. \\
%
%Several studies have been carried out in order to evaluate the impact and role that organized public health measures have in the midst of real epidemics \cite{Hatchett07-1,bootsma07-1,markel07-1},
%but only a few recent attempts have considered spontaneous social distancing phenomena. 
%
{Several studies have been carried out in order to evaluate the impact and role that organized public health measures have in the midst of real epidemics \cite{Hatchett07-1,bootsma07-1,markel07-1}. However, only a few recent attempts have considered self-induced behavioral changes individuals adopt during an outbreak in order to reduce the risk of infection.
}
In some approaches individual behaviors were modeled by modifying contact rates in response to the state of the disease~\cite{Fenichel2011, Chen2011, bootsma07-1, Bagnoli2007, goffman64-1}. In others new compartments representing individual responses were proposed~\cite{Poletti2011, poletti09-1, epstein08-1}. Finally, in some studies the spread of information in the presence of the disease was explicitly modeled and coupled with the spreading of the disease itself \cite{funk09-1}. However, we are still without a formulation of a general behavior-disease model.

In this study we propose a general framework to model the spread of information concerning the epidemic and the eventual behavioral changes in a single population. 
{The emergent infectious diseases that we consider throughout the manuscript resemble the natural history of an acute respiratory infection with a short duration of infectiousness and have mild impact on the health status of individuals in that healthy status is recovered at the end of the infectious period.}
We modify
 the classic susceptible-infected-recovered (SIR) model \cite{kermac27-1} by introducing a class of individuals, $S^{F}$, that represents susceptible people who self-initiate behavioral changes that lead to a reduction in the transmissibility of the infectious disease. In other words, this class models the spread of  `fear'  associated with the actual infectious disease spread
 \cite{epstein08-1, lynch91-1}. Individuals who fear the disease self-initiate social distancing measures that reduce the transmissibility of the disease. The spread of fear depends on the source and type of information to which individuals are exposed ~\cite{funk09-1,defleur89-1}. We classify the general interaction schemes governing the transitions of individuals into and out of $S^{F}$ by considering behavioral changes due to different information spreading mechanisms, i.e., belief-based versus prevalence-based and local versus global information spreading mechanisms.
 We provide a theoretical and numerical analysis of the various mechanisms involved and uncover a rich phenomenology of the behavior-disease models that includes epidemics with multiple activity peaks and transition points.
 We also show that in the presence of belief-based propagation mechanisms the population may acquire a collective `memory' of the fear of the disease that makes the population more resilient to future
 outbreaks. This abundance of different dynamical behaviors clearly shows the importance of the behavior-disease perspective in the study of  realistic progressions of infectious diseases and provides a chart for future studies and scenario analyses in data-driven epidemic models.

\section*{Methods}

\subsection*{Epidemic model and basic assumptions}
\label{model_1}
\begin{figure}
\begin{centering}
\includegraphics[width=0.4\textwidth]{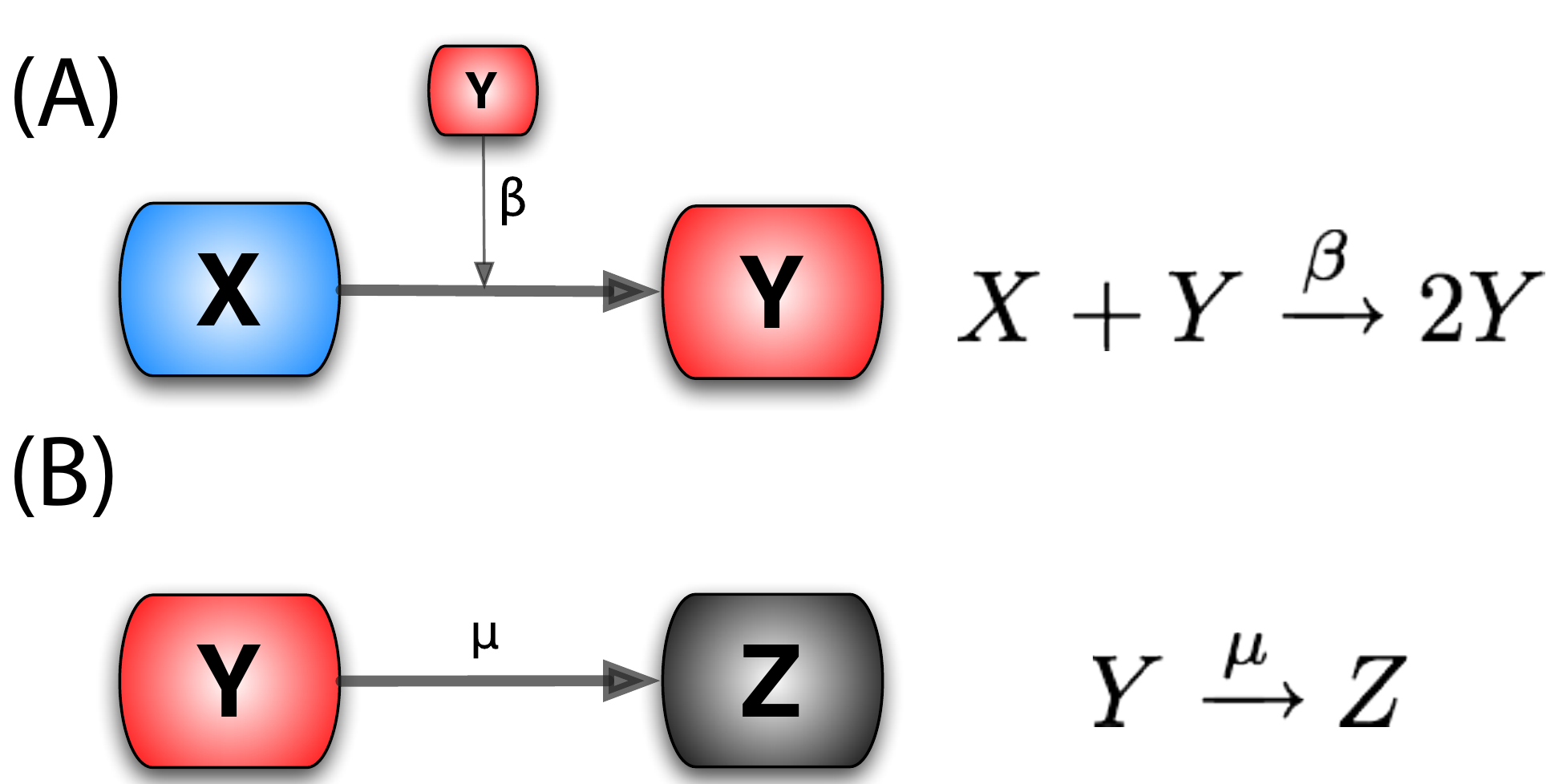}
\par\end{centering}

\caption{\label{trans} {\small Schematic representation of the two types of transitions that will be recurrent in the paper. In panel (A) we show the first in which individuals in compartment $X$ interact with individuals in class $Y$, represented by the small square,  becoming $Y$ themselves. 
In general the compartment inducing the transition of individuals in $X$ could be any other compartment in the model, e.g. $M$, 
different from the end-point of the transition. 
We assume the homogeneous mixing of the population so that the rate at which an individual in $X$ interacts with individuals in 
$Y$ and changes status is simply given by the product of prevalence $Y/N$ of $Y$ and the transmission rate $\beta$, $\beta Y/N$.
This type of reaction can be written as $X+Y \xrightarrow{\beta} 2Y$. In the case of the SIR model $X=S$ and $Y=I$.
 In panel (B) we show the second type. This is a spontaneous 
transition with rate $\mu$ in which an individual in  compartment $Y$ spontaneously moves to  compartment $Z$.
These types of reactions can be written as $Y \xrightarrow{\mu} Z$. In the SIR model $Y=I$ and $Z=R$.}}
\end{figure}

 In order to describe the infectious disease progression we use the minimal and prototypical SIR model. This model is customarily used to describe the progression of acute infectious diseases such as influenza in closed populations where the
total number of individuals 
$N$ in the population is partitioned into the compartments $S(t)$, $I(t)$ and $R(t)$, denoting  the number of susceptible, infected and recovered individuals at
time $t$, respectively. By definition it follows $N=S(t)+I(t)+R(t)$. The model is described by two simple types of transitions represented in Figure~(\ref{trans}).
The first one, denoted by $S\to I$, is when a susceptible individual interacts with an infectious individual and acquires infection with transmission rate $\beta$. 
The second one, denoted by $I\to R$, occurs when an infected individual recovers from the disease with rate $\mu$ and is henceforth assumed to have permanent immunity to the disease.
  The SIR model is therefore described by the two following reactions and the associated rates:
\begin{equation}
S+I \xrightarrow{\beta} 2I, 
\end{equation}
\begin{equation}
I \xrightarrow{\mu} R.
\end{equation}
While the $I\to R$ transition is itself a spontaneous process, the transition from $S\to I$ depends on the structure of the population and the contact patterns of individuals.
 Here we consider the usual homogeneous mixing approximation that
assumes that individuals interact randomly among the population. According to this assumption the larger the
number of infectious individuals among one individual's contacts the higher the probability of transmission of the infection.  This readily translates in the definition of the force of infection in terms of 
a mass action law~\cite{Keeling_Rohani}, $\lambda_{S\to I}=\beta I(t)/N$ that expresses the per capita rate at which susceptible individuals contract the infection. 
In order to simulate the SIR model as a stochastic process we can consider a simple binomial model of transition for discrete individuals and discrete times. 
{
Each member of the susceptible compartment at time $t$ has a probability $\lambda_{S\to I} \Delta t=\beta \Delta t I(t)/N$ during the time interval between $t$ and $t+\Delta t$ to contract the disease and transfer to the infected state at time $t+\Delta t$, where $\Delta t$ is the unitary time scale considered that we have set to $\Delta t= 1\;day$ in simulations.
As we assume to have $S(t)$ independent events occurring with the same probability, the number of newly infected individuals $I_+$ generated during the time interval $\Delta t$ is a random variable that will follow the binomial distribution $Bin \left [ S(t), \beta \Delta t I(t)/N \right]$.  
Analogously, the number of newly recovered individuals $R_+$ at time $t+\Delta t$ is a random variable that will obey the binomial distribution $Bin \left[ I(t), \mu \Delta t \right]$, where the number of independent trials is given by the number of infectious individuals $I(t)$ that attempt to recover and the probability of recovery in the time interval $\Delta t$ is given by the recovery probability $\mu \Delta t$.
In this processes we recognize that the stochastic variables define a Markov chain~\cite{Longini98,Halloran10} of stochastic events  $\{ S(t), I(t), R(t) : t=0,\Delta t, 2\Delta t, \ldots\}$ in which the current state of the system is
 determined only by the state of the system at the previous time steps. Formally, we can indeed write the following Markov chain relations:
 }
\begin{eqnarray}\nonumber
S(t+\Delta t) &=& S(t) - I_+, \\
I(t+\Delta t) &=&I(t) + I_+ - R_+, \\ \nonumber
R(t+\Delta t) &=&R(t) + R_+.
\end{eqnarray}
These equations can be readily used to simulate different stochastic realizations of the epidemic events with the same basic parameters and initial conditions. This allows us to analyze the model's behavior by taking into account statistical fluctuations and noise in the epidemic process.
The equations can also be  translated into the standard set of continuous deterministic differential equations describing the SIR model by using expected values as
\begin{eqnarray*}
d_{t}S(t) & = & -\beta S(t)\frac{I(t)}{N},\\
d_{t}I(t) & = & -\mu I(t)+\beta S(t)\frac{I(t)}{N},\\
d_{t}R(t) & = & \mu I(t).
\end{eqnarray*}
The crucial parameter in the analysis of single population epidemic
outbreaks is the basic reproductive 
number $R_0$, which counts the expected number of secondary infected cases 
generated by a primary infected individual.
Under the assumption of homogeneous mixing of the population 
the basic reproductive number of the SIR model is given by 
\begin{equation}
  R_0 = \frac{\beta}{\mu}.
\label{eq:R0}
\end{equation}
By the simple linearization of the above equations for $I/N \ll 1$ it is straightforward to see that in the single
population case any epidemic will spread to a nonzero fraction of the population only if $R_0>1$.
In this case the epidemic is able to generate a number of infected
individuals larger than those who recover,  leading to an
increase in the overall number of  infectious individuals $I(t)$. 
The previous considerations lead to the definition of a crucial
epidemiological concept: the epidemic threshold. Indeed, if the
transmission rate is not large enough  to allow a reproductive number
larger than one (i.e., $\beta>\mu$), the epidemic outbreak will be confined to a tiny portion of the population and will die out in a finite
amount of time in the thermodynamic limit of $N \rightarrow \infty$. 

In the following we will use binomial stochastic processes to simulate numerically the progression of the epidemics and we will use the continuum limit to provide the analytical discussion of the models.

\subsection*{Coupling epidemic spreading and behavioral changes}
We need to classify the source and type of information concerning the disease that people use to conduct their behavior in order to model the coupling between behavioral changes and the disease spread.
 In other words, while the disease spreads in the population, individuals are exposed (by local contacts, global mass media news, etc.) to information ~\cite{funk09-1} on the disease that will lead to changes in their behavior. This is equivalent to the coupled spread of two competing contagion processes~\cite{funk09-1, poletti09-1, epstein08-1}: the infectious disease  and the `fear of the disease' contagion processes.
The fear of the disease is what induces  behavioral changes in the population. For this reason we will assume that individuals affected by the fear of the disease will be grouped in a specific compartment $S^{F}$ 
of susceptible individuals. These individuals will not be removed from the population, but they will take actions such as reducing the number of potentially infectious contacts, wearing face masks, and other social distancing measures that change  disease parameters. 
In the following we will consider that self-induced behavior changes have the effect of reducing the transmission rate of the infection, introducing the following reaction:
\begin{equation}
S^F+I \xrightarrow{r_\beta \beta} 2I, 
\end{equation}
with $ 0 \leq r_\beta<1 $ (i.e., $r_{\beta}\beta< \beta$). 
The above process corresponds to a force of infection on the individuals affected by the fear contagion $\lambda_{S^F\to I}=r_{\beta}\beta I(t)/N$. The parameter $r_\beta$
 therefore modulates the level of self-induced behavioral change that leads to the reduction of the transmission rate. 
As the scope of the awareness of the disease or of the adopted behavioral changes is avoidance of infection, we assume that individuals in the $S^F$ compartment relax their behavioral changes upon infection and return back to their regular social behavior. 
While the above modeling scheme is a straightforward way to include
 social distancing in the system,  a large number of possible scenarios can be considered in the modeling of the contagion process that induce susceptible individuals to adopt self-induced behavioral changes and transition to the state $S^F$.
In particular we consider three main mechanisms:
\begin{itemize}
\item {\bf Local, prevalence-based spread of the fear of the disease}. In this scenario we assume that susceptible individuals will adopt behavioral changes only if they interact with infectious individuals. This implies that the larger the
number of sick and infectious individuals among one individual's contacts, the higher the probability for the individual to adopt behavioral changes induced by  awareness/fear of the disease.
 The fear contagion process therefore can be  modeled as
\begin{equation}
S+I \xrightarrow{\beta_F} S^F + I,
\end{equation}
where in analogy with the disease spread, $\beta_F$ is the transmission rate of the awareness/fear of the disease.
  This process defines a transition rate for the fear of the disease that can be expressed by the usual mass-action law $\lambda^{I}_{S\to S_F}=\beta_F I(t)/N$.
\item {\bf Global, prevalence-based spread of the fear of the disease}. In some circumstances, individuals adopt self-induced behavioral changes
 because of information that is publicly available, generally through newspapers, television, and the Internet. In this case the local transmission is superseded by a global
 mechanism in which the news of a few infected individuals, even if not in contact with the large majority of the population, is able to trigger a widespread reaction in
 the population. In this case the fear contagion process is not well represented by the usual mass action law and has to be replaced by
\begin{equation}
\beta_{F} \frac{I(t)}{N}\rightarrow \lambda^{II}_{S \rightarrow S^F}=\beta_{F}(1-e^{-\delta I(t)}),
\end{equation}
where $0<\delta\leq1$. 
Figure~(\ref{trans3}) shows the schematic representation of this.

\begin{figure}
\begin{centering}
\includegraphics[width=0.3\textwidth]{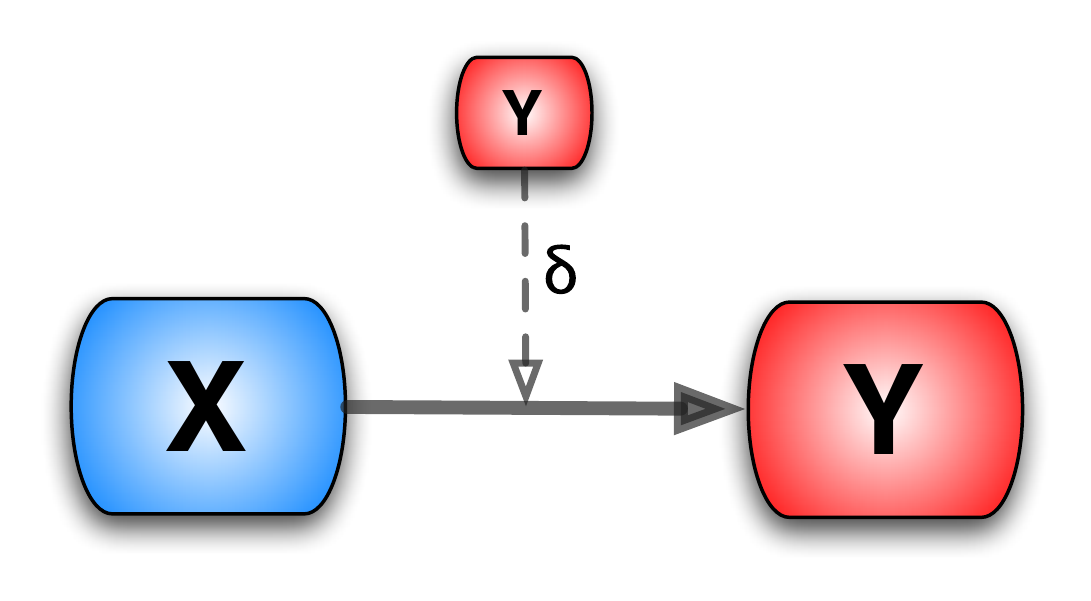}
\par\end{centering}
\caption{\label{trans3} {\small Schematic representation of the third type of interaction discussed. In this case the transition into  compartment $Y$ is based on the absolute number of the individuals in the 
compartment (shown by the small square). In general the inducing compartment could be different (e.g. $M$) than the end-point of the transition. }}
\end{figure}

For small values of $\delta$ we have a pseudo mass action law~\cite{Keeling_Rohani} of the first order in $\delta$:
\begin{equation}
\beta_{F}(1-e^{-\delta I(t)})=\beta_{F}\left[\delta I(t)+\mathcal{O}(\delta^{2})\right].
\end{equation}
The above contagion process acts on the whole population even in the case of a very limited number of infectious individuals and the parameter $\delta^{-1}$ identifies the characteristic number of infected individuals reported by the news above which the  fear spreads quickly in the population similarly to a panic wave.  
\item {\bf Local, belief-based spread of the fear of the disease}. In addition to the local prevalence-based spread of the fear of the disease, in this case we assume that the fear contagion may also occur by contacting individuals who have already acquired  fear/awareness of the disease.
 In other words, the larger the
number of individuals who have fear/awareness of the disease among one individual's contacts, the higher the probability of that individual adopting behavioral changes and moving into the $S^F$ class.
 The fear contagion  therefore can also progress according to the following process:
\begin{equation}
S+S^{F}\xrightarrow{\alpha\beta_{F}}2S^{F},
\end{equation}
where the transmission rate is $\alpha \beta_{F}$, with $\alpha$ modulating the ratio between the transmission rate by contacting infected individuals and contacting individuals with  fear of the disease. The 
transition rate is defined by 
the mass-action law $\lambda^{III}_{S \rightarrow S^F} = \alpha \beta_F  S^F(t)/N$.
\end{itemize}
The fear/awareness contagion process is not only defined by the spreading of  fear from individual to individual, but also by the process defining the transition from the state of
 fear of the disease back to the regular susceptible state in which the individual relaxes the adopted behavioral changes and returns to  regular social behavior.  We can imagine a similar
 reaction $S^F\to S$ on a very long time scale in which individuals lose memory of the disease independent of their interactions with other individuals and resume their normal social behavior. This would correspond to spontaneous recovery from fear as proposed by Epstein {\it et al.} \cite{epstein08-1}.
 However, our social behavior is modified by our local interactions with other individuals  on a much more rapidly acting time-scale. We can therefore consider the following processes:
\begin{equation}
S^F+S \xrightarrow{\mu_F} 2S
\end{equation} 
and
\begin{equation}
S^F+R \xrightarrow{\mu_F} S + R.
\end{equation}
We can then define two mass-action laws: $\lambda^{A}_{S^F \rightarrow S}=\mu_F S(t)/N$ and $\lambda^{B}_{S^F \rightarrow S}=\mu_F R(t)/N$. These mimic
 the process in which the interaction between individuals with fear and without fear, susceptible or recovered, leads the individual with fear to resume regular social behavior. Both processes, occurring with rate $\mu_F$, tell us that the larger the
number of individuals who adopt regular social behavior among one individual's contacts, the higher the probability for the individual to relax any
 behavioral changes and resume regular social behavior. 
 The two interactions translate into a unique mass action law: $\lambda^{A}_{S^F\to S}+\lambda^{B}_{S^F\to S}=\lambda_{S^F\to S}=\mu_F (S(t)+R(t))/N$. The fear contagion process is therefore hampered by the presence of large numbers of individuals acting normally. The  spreading of fear is the outcome of two opposite forces acting on  society, 
but is always initially triggered by the presence of infectious individuals~\cite{funk09-1, poletti09-1, epstein08-1, Bagnoli2007, bootsma07-1}. 
In Table~\ref{tab1} we report all the infection and recovery transitions for the disease and fear contagion dynamics and the corresponding terms and rates. We will use those terms  to characterize different
 scenarios of  interplay between the information and disease spreading processes. Unless specified otherwise the numerical simulations will be performed by individual-based chain binomial
 processes~\cite{Longini98,Halloran10} in discrete time and the analytical discussion will consider the continuous deterministic limit. In the comparison between the analytic conclusions with the numerical simulations we will
 always make sure to discuss the differences due to stochastic effects such as the outbreak probability at relatively small values of the reproductive number $R_0$. In the following discussion  $R_0$ will refer to the basic reproductive number of the SIR model unless specified otherwise.

\begin{table*}
 \renewcommand\multirowsetup{\centering}
\begin{tabular}{llccl}
\hline 
&\textbf{Transition}  & \textbf{Transition rate} & \textbf{Equation flow term} & \textbf{Dynamical model} \tabularnewline [0.5ex]
 \cline{2-5}
\multirow{3} {*}{ \begin{sideways}\centering \bf  Disease \end{sideways} } 
&$S+I \xrightarrow{\beta} 2I$ & $\lambda_{S \rightarrow I}=\beta \frac{ I(t)}{N}$ & $\beta \frac{ I(t)}{N} S(t)$ &  Models I,II,III \tabularnewline [3ex]
&$I \xrightarrow{\mu} R$& $\mu$&  $\mu I(t)$ & Models I,II,III \tabularnewline [0.5ex]
 \cline{2-5} 
\multirow{6} {*} { \begin{sideways} \bf Behavior \end{sideways}} 
&$S^F+I \xrightarrow{r_\beta \beta} 2I$ & $\lambda_{S^F\rightarrow I} = r_\beta \beta \frac{ I(t)}{N}$ &  $r_\beta \beta \frac{ I(t)}{N}S^F(t)$ &Models I,II,III \tabularnewline [1.5ex]
&$S+I \xrightarrow{\beta_F} S^F+I$ & $\lambda^{I}_{S\rightarrow S^F}=\beta_F\frac{ I(t)}{N}$ & $\beta_F\frac{ I(t)}{N} S(t)$ &Models I,II,III \tabularnewline [1.5ex]
&$S+I \xrightarrow{\beta_F} S^F+I$ & $\lambda^{II}_{S\rightarrow S^F} = \beta_F \left[ 1-e^{-\delta I(t)} \right] $ &  $\beta_F \left[ 1-e^{-\delta I(t)} \right]S(t) $ & Model II \tabularnewline [1.5ex]
&$S+S^F \xrightarrow{\beta_F \alpha } 2S^F$ & $\lambda^{III}_{S\rightarrow S^F} = \beta_F \alpha \frac{S^F(t)}{N}$ &  $\beta_F \alpha \frac{S^F(t)}{N}S(t)$ & Model III \tabularnewline [1.5ex]
&$S^F+S \xrightarrow{\mu_F} 2S$ & $\lambda^{A}_{S^F\rightarrow S} =\mu_F \frac{S(t)}{N}$ &  $\mu_F \frac{S(t)}{N}S^F(t)$ & Models I,II,III \tabularnewline [1.5ex]
&$S^F+R \xrightarrow{\mu_F} S+R$ & $\lambda^{B}_{S^F\rightarrow S} = \mu_F \frac{R(t)}{N}$&  $\mu_F \frac{R(t)}{N}S^F(t)$ &Models I,II,III  \tabularnewline  [0.5ex]
\hline
\end{tabular}
\caption{\label{tab1} In this table we show all the transitions and their rates used in the three different models. In the last column we write the model in which the transition has been used.
For example, the first transition $S+I \xrightarrow{\beta} 2I$ is related to the disease dynamic and has been used in all three models. }
\end{table*}

\section*{Results}
\subsection*{Model I: Local, prevalence-based spread of the fear of the disease}
\label{model_11}
The first model (Model I) we consider is the coupling of the SIR model with local prevalence-based spread of the fear of the disease.
The coupled behavior-disease model is  described by the following set of
equations:
\begin{widetext}
\begin{eqnarray}
d_{t}S(t) & = & -\lambda_{S \rightarrow I}S(t)- \lambda^I_{S \rightarrow S^F}S(t)+\lambda_{S^F \rightarrow S}S^{F}(t),\nonumber\\
d_{t}S^{F}(t) & = & -\lambda_{S^F \rightarrow I}S^{F}(t)+ \lambda^I_{S \rightarrow S^F}S(t)-\lambda_{S^F \rightarrow S}S^{F}(t),\\
d_{t}I(t) & = & -\mu I(t)+\lambda_{S \rightarrow I}S(t)+ \lambda_{S^F \rightarrow I}S^F(t),\nonumber\\
d_{t}R(t) & = & \mu I(t).\nonumber
\end{eqnarray}
\end{widetext}

\begin{figure}
\begin{centering}
\includegraphics[width=0.5\textwidth]{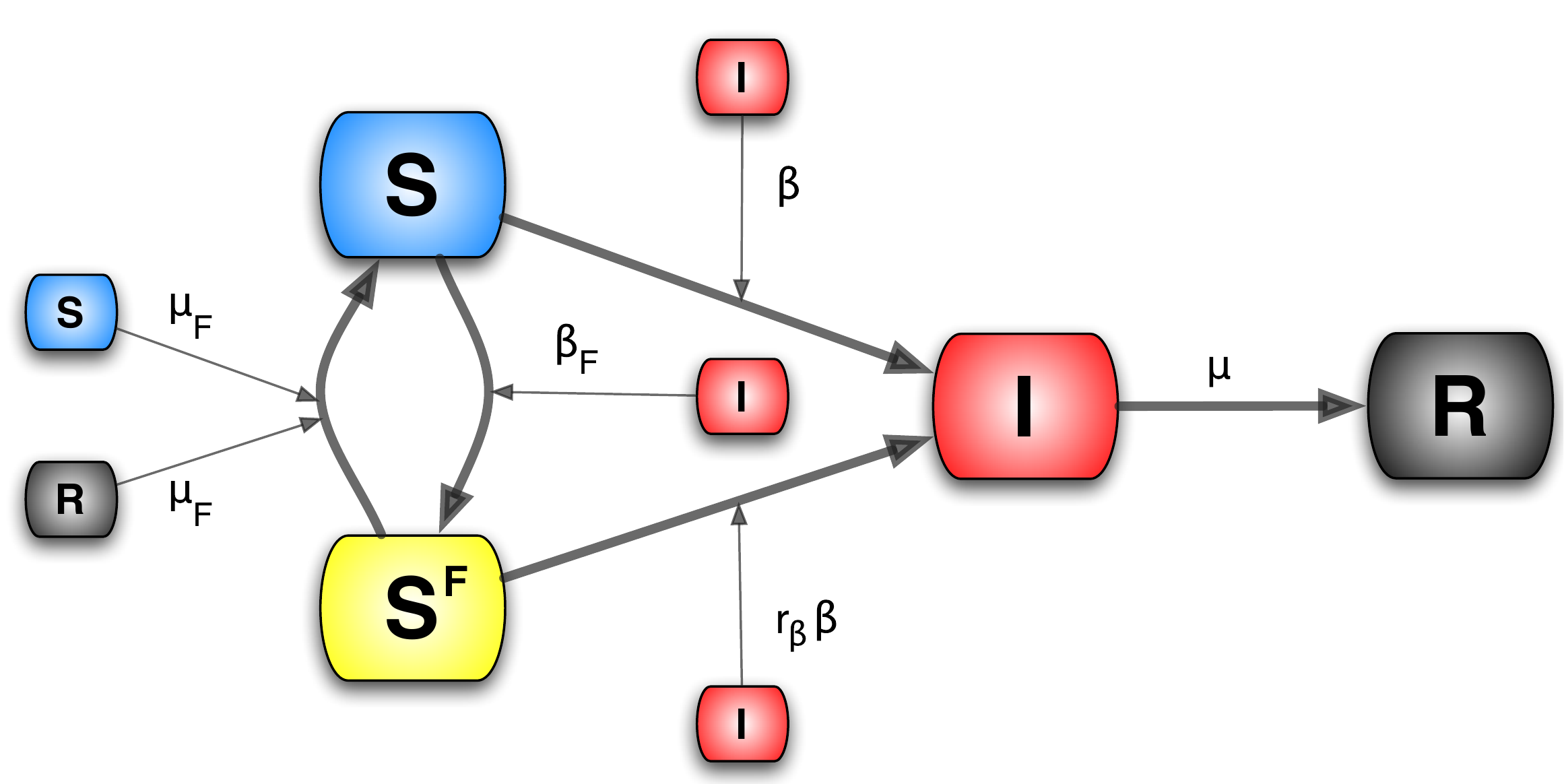}
\par\end{centering}
\caption{\label{modelI} {\small {\bf Model I.} Schematic representation of Model I. }}
\end{figure}

A schematic representation of the model is provided in Figure~(\ref{modelI}). 
Considering Table~\ref{tab1} we can write down all the terms,
\begin{widetext}
\begin{eqnarray}
\label{f_o_f}
d_{t}S(t) & = & -\beta S(t)\frac{I(t)}{N}-\beta_{F}S(t)\frac{I(t)}{N}+\mu_{F}S^{F}(t)\left[\frac{S(t)+R(t)}{N}\right],\\ \nonumber
d_{t}S^{F}(t) & = & -r_{\beta}\beta S^{F}(t)\frac{I(t)}{N}+\beta_{F}S(t)\frac{I(t)}{N}-\mu_{F}S^{F}(t)\left[\frac{S(t)+R(t)}{N}\right],\\ \nonumber
d_{t}I(t) & = & -\mu I(t)+\beta S(t)\frac{I(t)}{N}+r_{\beta}\beta S^{F}(t)\frac{I(t)}{N},\\ \nonumber
d_{t}R(t) & = & \mu I(t), 
\end{eqnarray}
\end{widetext}
% \begin{eqnarray}
% \label{f_o_f}
% d_{t}S(t) & = & -\beta S(t)\frac{I(t)}{N}-\beta_{F}S(t)\frac{I(t)}{N}+\\ \nonumber
% &+&\mu_{F}S^{F}(t)\left[\frac{S(t)+R(t)}{N}\right],\\ \nonumber
% d_{t}S^{F}(t) & = & -r_{\beta}\beta S^{F}(t)\frac{I(t)}{N}+\beta_{F}S(t)\frac{I(t)}{N}-\\ \nonumber
%  &-&\mu_{F}S^{F}(t)\left[\frac{S(t)+R(t)}{N}\right],\\ \nonumber
% d_{t}I(t) & = & -\mu I(t)+\beta S(t)\frac{I(t)}{N}+r_{\beta}\beta S^{F}(t)\frac{I(t)}{N},\\ \nonumber
% d_{t}R(t) & = & \mu I(t), 
% \end{eqnarray}
in which
\begin{equation}
\sum_{i}d_{t}X_{i}(t)=0 \;\; {\rm for} \,\ \forall \,\ t \;\; {\rm and} \,\ X_{i} \in \left[ S,S^{F},I,R \right],
\end{equation}
meaning that the total number of individuals in the population does not change.
 In acute diseases, the time scale of the  spreading is very small with respect to the  average lifetime of a person, allowing us to ignore  birth and death processes and the demographic drift of the population.
 This is also the time scale over which it is more meaningful to consider the effect of the spread of behavioral changes. Diseases with a longer time scale may be equally affected by behavioral changes emerging especially as cultural changes toward certain social behavior -- for instance sexual habits in the presence of a sexually transmitted disease with a long latency period -- but in this case the demography of the system should be taken into account.

To explain the equations we can simply consider the negative terms. In particular the first term of the first equation
 in Eq. (\ref{f_o_f}) takes into account individuals in the susceptible compartment $S$ who through interaction 
with infected individuals become sick.
The second term takes into account individuals in the susceptible compartment $S$ who through interaction with infected individuals 
change their own behavior. The first term of the second equation takes into account individuals in  compartment $S^F$ who through interaction with infected individuals become sick. It is important to remember that the 
transmission rate for people in  compartment $S^F$ is reduced by a factor $r_\beta$ due to the protection that they gain on account of membership in this class. The last term in the second equation takes into account people in  compartment $S^F$ 
who through interaction with healthy individuals, $S$, and recovered ones, $R$, normalize their behavior and move back to  compartment $S$.
The first term in the third equation takes into account the spontaneous recovery of sick individuals.

\begin{figure}
\begin{centering}
\includegraphics[width=0.5\textwidth]{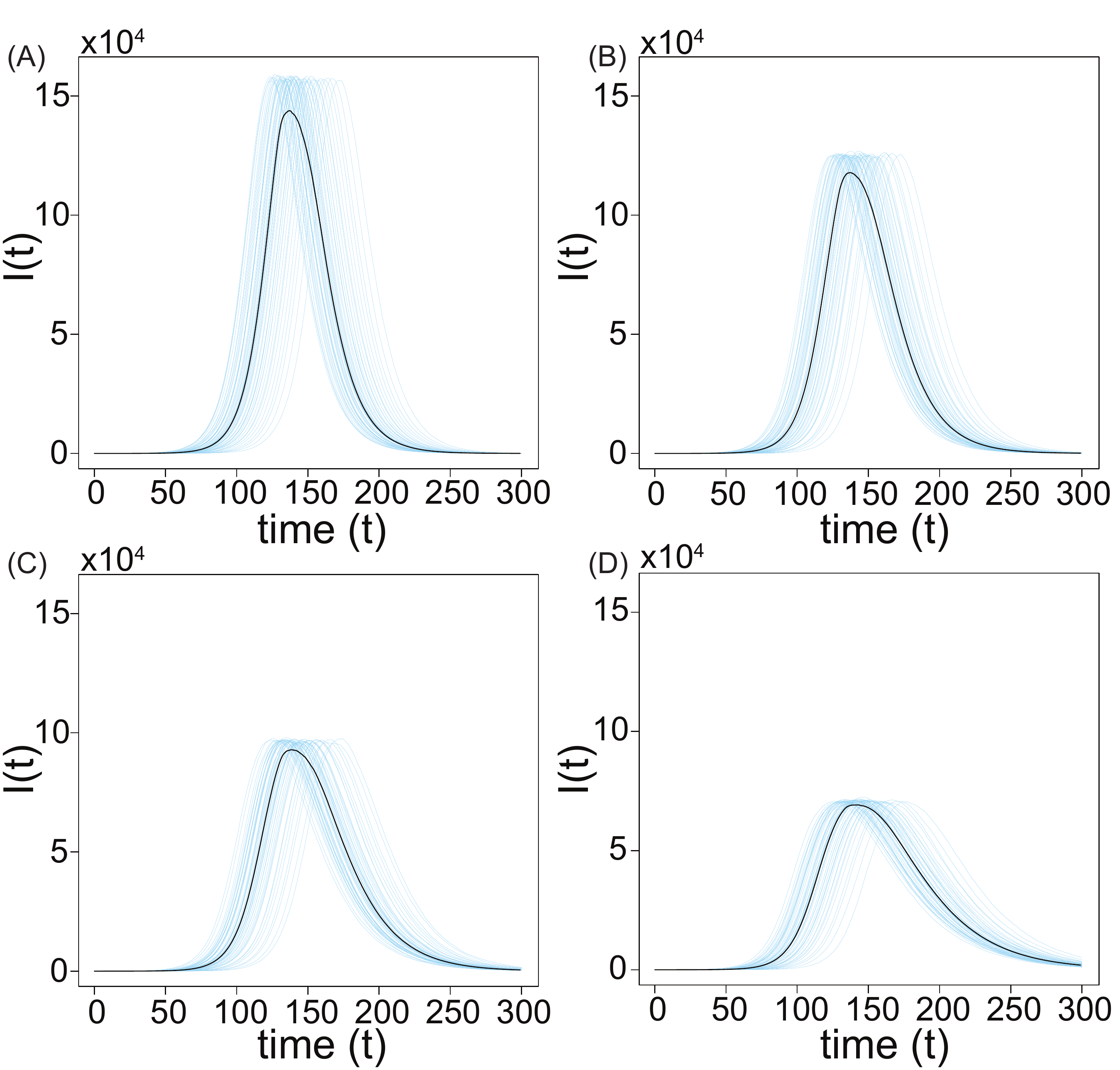}
\par\end{centering}

\caption{\label{curves} {\small {\bf Model I} for $\mu_{F}=0.5\;day^{-1}$, $r_{\beta}=0.5$, $\mu=0.1\;day^{-1}$, $N=10^6$ and $R_0=2$. We show the medians
of $I(t)$, evaluated using $5 \times 10^3$ stochastic runs for the baseline (SIR model without fear of contagion) and three realizations of the model for different values of $\beta_F$. In particular in panel (A) we show the baseline SIR model with the 
same disease parameters. In panel (B) we set $\beta_F=1\;day^{-1}$. In panel (C) we set $\beta_F=2.5\;day^{-1}$. In panel (D) we set $\beta_F=5\;day^{-1}$. It is clear how the peak time is the same for all the scenarios and how the number of 
infected individuals at peak is reduced as $\beta_F$ increases. }}

\end{figure}

\begin{figure}
\begin{centering}
\includegraphics[width=0.5\textwidth]{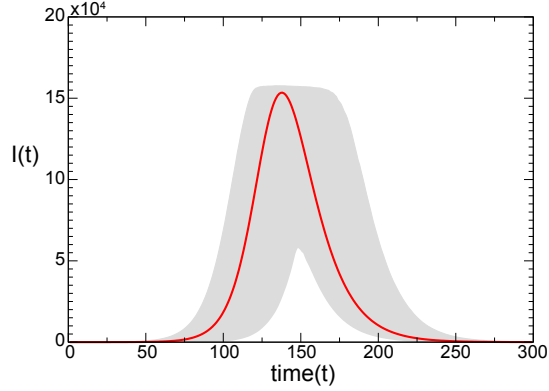}
\par\end{centering}

\caption{\label{comparison} {\small {\bf Model I} fixing $\beta_{F}=0.25\;day^{-1}$, $r_{\beta}=0.5$, $\mu=0.1\;day^{-1}$ and $R_0=2$. We compare the solution of the deterministic equations (red solid line) with the $95 \%$ reference ranges 
of our stochastic solutions. Here we consider $5 \times 10^3$ runs that produced at least an epidemic size of $0.1 \%$ of the population ($N=10^6$).}}

\end{figure}

 It is natural to assume that in the beginning of 
the disease spreading process the population is fully susceptible except for the infectious seeds, which means that we can set $S^F(t=0)=0$. At this point the behavioral response is not active yet. 
If the disease proceeds to spread much faster than fear contagion, then the model reduces to the classic SIR with basic reproductive number $R_{0}=\beta/\mu$. In this case the initial spread is well described
by $I(t \sim 0) \sim S^F(t \sim 0) \sim R(t \sim 0)\sim 0$. The number of individuals in the compartment $S^F$ is of the same order of infectious and recovered individuals. From the conservation of 
the number of individuals follows $S(t \rightarrow 0) \sim N$. Since $S$ is the leading order, all the terms in the equations like $X(t)Y(t)$ in which both $X$ and $Y$ are different from $S$
 can be considered as second order. Using this approximation we can linearize the system and reduce the equations to first-order ordinary differential equations that are easy to integrate. In particular for $S^F$ 
we can write
\begin{equation}
d_t S^F(t)=+\beta_F I(t)-\mu_F S^F(t),
\end{equation}
which has the following solution:
 \begin{equation}
\label{sol_early_m_I}
S^{F}(t)\sim\frac{\beta_{F}}{\mu(R_{0}-1)+\mu_{F}}\left(e^{\mu(R_{0}-1)t}-e^{-\mu_{F}t}\right).
\end{equation}
For $R_0>1$ fear will spread in the population since the condition $\mu (R_0-1)> -\mu_F$ is always satisfied. The growth of the fear contagion is due to the spread of the infection in the population.\\
When fear spreads much faster than the disease, $\beta_{F} \gg \beta$,
everyone quickly becomes scared and our model reduces to an SIR
model with a reduced reproductive ratio $R^F_{0}=r_{\beta}\beta/\mu=r_\beta R_0$
that is dominated by the characteristics of the $S^{F}$ compartment.
\\
By considering both stochastic simulations of the model and direct integration of the equations, we explored numerically the intermediate regime between these two
limits, i.e. $\beta_{F}/\beta \sim \mathcal{O}(1)$. The spread of the fear of infection contagion in this regime does not
significantly affect the timing of the disease spread, as showed in Figure~(\ref{curves}). 
In this figure the stochastic fluctuations are demonstrated by $50$ individual realizations and compared with the median profiles obtained by considering $5 \times 10^3$ different stochastic realizations. 
The deterministic solution of the equation for $I(t)$, obtained by direct integration
 of the equations, is well inside the $95 \%$ reference range of our stochastic simulations as shown in Figure~(\ref{comparison}).  In this region of the model's phase space   fear simply produces a mild reduction in the epidemic size.

% \begin{figure}
% \begin{centering}
% \includegraphics[width=0.5\textwidth]{fig/Untitled-1.pdf}
% \par\end{centering}
% 
% \caption{\label{r_rf0_mufs} {\small {\bf Model I:}  epidemic size $R_{\infty}/N$ for different
% values of $\mu_{F}$, $r_{\beta}$ fixing $\beta_{F}=0.25$, $\mu=0.1$ and $R_0=2$. The surface in the phase space has been determined through direct numerical integration of the equations.}}
% 
% \end{figure}

\begin{figure}
\begin{centering}
\includegraphics[width=0.4\textwidth]{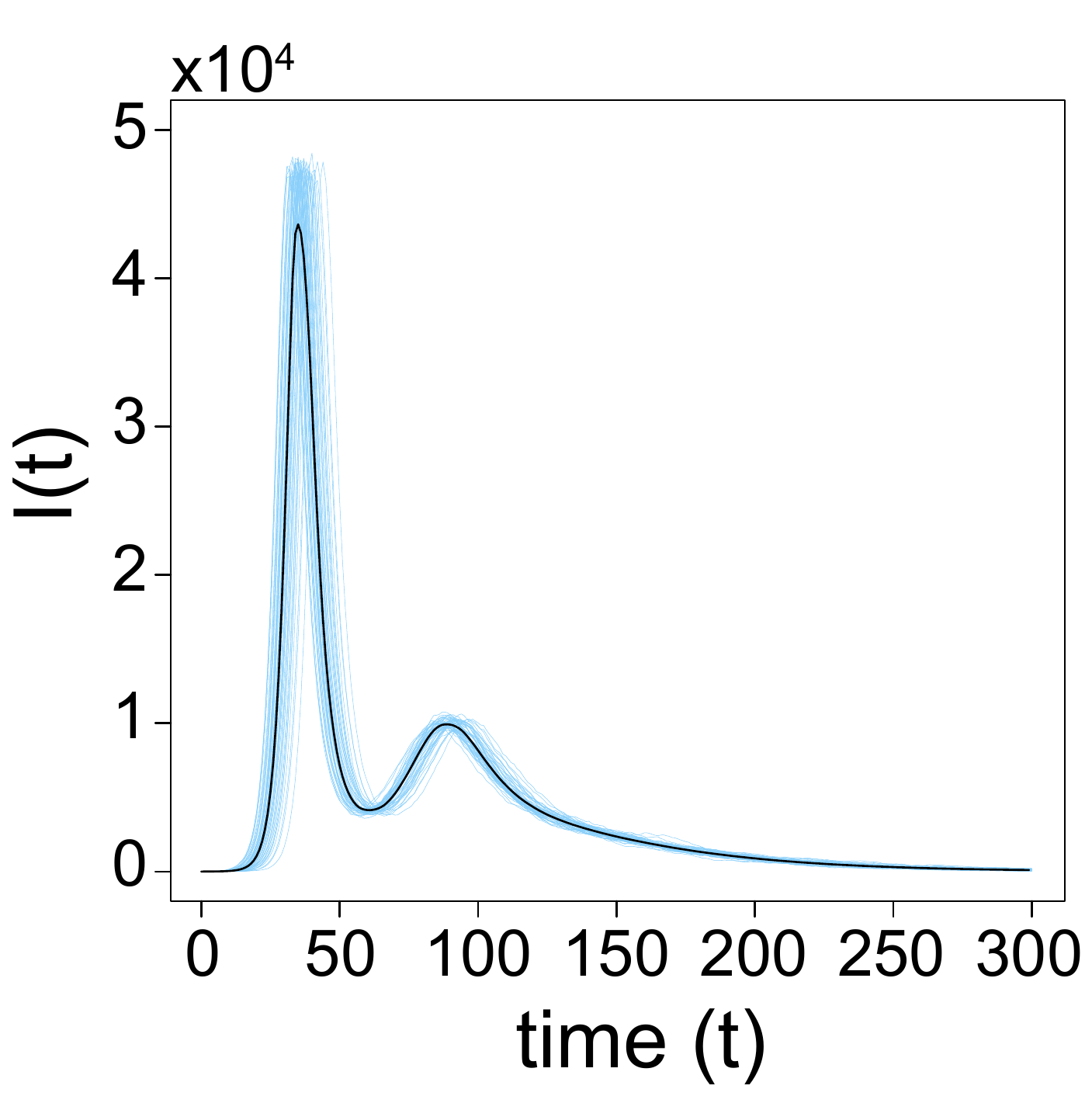}
\par\end{centering}

\caption{\label{two-peak_model1} {\small {\bf Model I} Multiple waves of infection. Fixing $\mu=0.1\;day^{-1}$, $R_0=2$, $\beta_F=3\;day^{-1}$, $\mu_F=0.1\;day^{-1}$, $N=10^6$ and $r_\beta =0.1$ we show $100$ stochastic runs of the infected profiles and the median evaluated 
considering $5 \times 10^3$ runs in which the epidemic size is at least $0.1 \% $ of the population.}}

\end{figure}

\begin{figure*}
\begin{centering}
\includegraphics[width=1.0\textwidth]{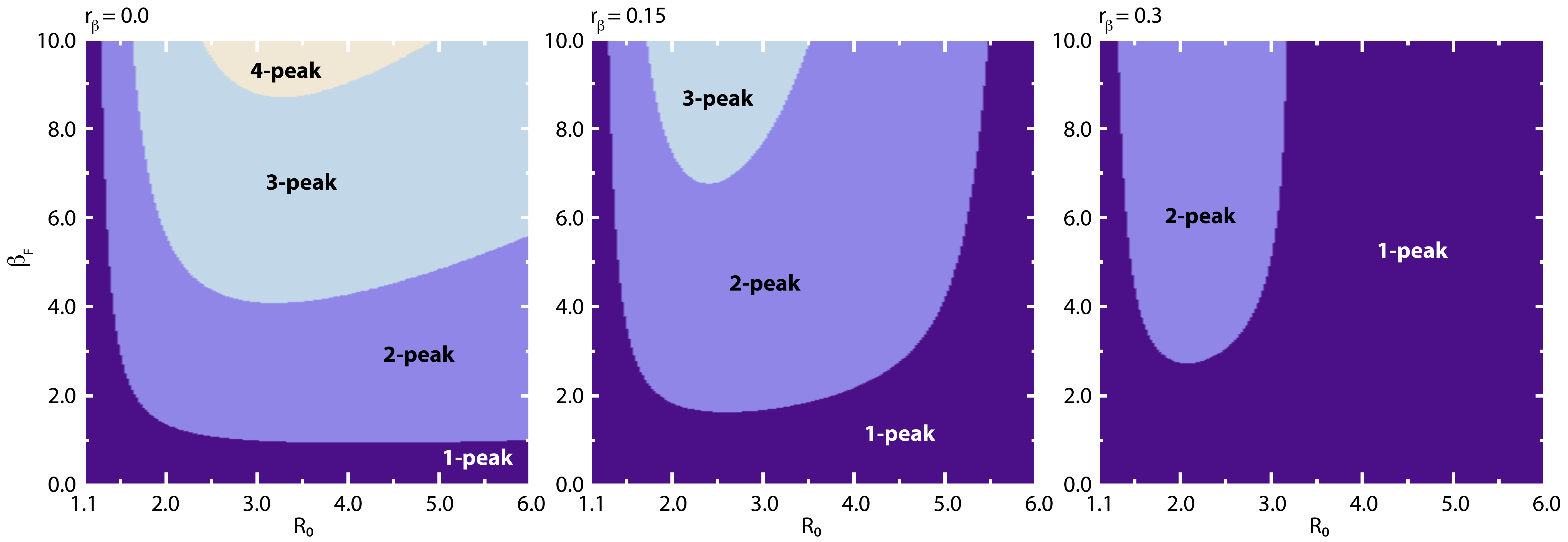}
\par\end{centering}

\caption{\label{multipeak_model1} {\small {\bf Model I} Phase diagram of infection waves on $R_0$-$\beta_F$ plane.
We display the regions of parameter space  on $R_0$-$\beta_F$ plane exhibiting different number of disease activity peaks for  three different values of $r_\beta=0, \; 0.15,\; 0.3$, where we have fixed $\mu=0.1\;day^{-1}$, $\mu_F=0.1\;day^{-1}$ and $N=10^6$. The phase diagram has been obtained by numerical integration of the deterministic equations in Eq.~\eqref{f_o_f}.
}}

\end{figure*}

\begin{figure}
\begin{centering}
\includegraphics[width=0.5\textwidth]{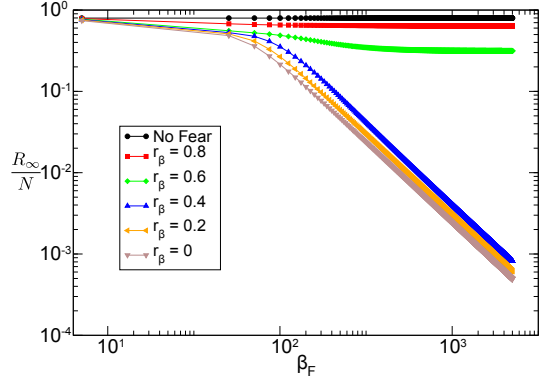}
\par\end{centering}

\caption{\label{rid_R_mod1} {\small {\bf Model I} fixing $\mu_{F}=0.5\;day^{-1}$, $\mu=0.1\;day^{-1}$, $N=10^6$ and $R_0=2$ we evaluate  the normalized epidemic size $R_\infty/N$ for different values
 of $\beta_F$ and 
$r_\beta$ through direct integration of the equations. Once the product $r_\beta R_0$ is smaller than unity, then the epidemic size goes to $0$ as $\beta_F \rightarrow \infty$.}}
\end{figure}

By increasing the value of $\beta_F$ it is possible to find a region of parameters characterized by multiple peaks. In Figure~(\ref{two-peak_model1}) we show $50$ stochastic runs and the median profile
obtained from $5 \times 10^3$ runs for a set of parameters associated with multiple peaks. After the first wave of infection individuals leave
 the compartment $S^F$ and return to the susceptible state in which they are less protected from the disease. The second wave manifests
 if the number of infected individuals at this stage is not too small and if there is still a large enough  pool of individuals susceptible to the infection.
{ 
A closer inspection of the parameter space by numerical integration of the deterministic equations yields very rich dynamical behavior. Figure~(\ref{multipeak_model1}) displays the phase diagram of the model on 
$R_0$-$\beta_F$ plane regarding different number of disease activity peaks for a set of model parameters. 
As $r_\beta$ increases, the region in which multiple peaks are encountered shifts to smaller values of $R_0$ and larger values of $\beta_F$. Fixing $r_\beta$, increasing values of $\beta_F$ increase the number of infection peaks while an increase in $R_0$ leads to a decrease in the number of peaks.
} 
It is interesting to note 
that adding a simple modification to the basic SIR model leads to scenarios with more than one peak. This is important not only from a mathematical point of view (existence of states characterized by 
multiple and unstable stationary points in the function $I(t)$) but also for practical reasons; in historical data from the 1918 pandemic multiple epidemic peaks were observed~\cite{bootsma07-1,Hatchett07-1, markel07-1}.
By increasing the value of $\beta_{F}$ to larger and larger values, the spread  of the fear contagion becomes increasingly rapid with respect to the spread of the disease. It is natural to think in this regime that the reproductive number 
of the disease is characterized by the $S^F$ class. We then have two different scenarios: 
\begin{enumerate}
\item If $r_{\beta}\beta/\mu>1$, then the epidemic size is given by that of an SIR model with $\beta\rightarrow\beta r_{\beta}$; 
\item If $r_{\beta}\beta/\mu<1$, then  fear completely stops the spreading of the disease.
\end{enumerate}
This is confirmed in Figure~(\ref{rid_R_mod1}) in which we plot the proportion of recovered individuals at the end of the epidemic, which is evaluated by the  integration of the deterministic equations. We consider
 different values of $\beta_F$ and $r_\beta$ and hold fixed the other parameters. 
It is clear that for very large values of $\beta_F$ the spreading of the disease is characterized by the reproductive number $r_\beta R_0$.\\
At the end of the disease epidemic the system enters the so-called `disease-free' stage. This region of the phase space is described by
\begin{equation}
(S,S^F,I,R )= ( S,S^F,0,R_{\infty} ).
\end{equation}
This regime can be easily derived by setting $I(t)=0$ in the set of Eqs.~(\ref{f_o_f}). The system is then reduced to
\begin{eqnarray}
d_{t}S(t) & = & \mu_{F}S^{F}(t)\left[\frac{S(t)+R(t)}{N}\right],\\ \nonumber
d_{t}S^{F}(t) & = & -\mu_{F}S^{F}(t)\left[\frac{S(t)+R(t)}{N}\right],\\ \nonumber
d_{t}I(t) & = & 0, \\ \nonumber 
d_{t}R(t) & = & 0.
\end{eqnarray}
From the last equation it is clear that $R(t)=constant=R_\infty$, and the first and second equations are  equivalent. It is then possible to find the solution for $S^F$ and $S$ by using the conservation of individuals. In particular the equation to solve is
\begin{eqnarray}
d_t S^F(t)&=&-\mu_{F}S^{F}(t)\left[\frac{S(t)+R_\infty}{N}\right]\nonumber\\
&=&-\mu_{F}S^{F}(t)\left[\frac{N-S^F(t)}{N}\right].
\end{eqnarray}
% \begin{eqnarray}
% d_t S^F(t)&=&-\mu_{F}S^{F}(t)\left[\frac{S(t)+R_\infty}{N}\right]\\ \nonumber 
% &=&-\mu_{F}S^{F}(t)\left[\frac{N-S^F(t)}{N}\right].
% \end{eqnarray}
By integrating this equation directly  it is easy to show that fear disappears exponentially:
\begin{equation}
S^{F}(t)\sim e^{-\mu_{F}t}.
\end{equation}
In the stationary state, for $t \rightarrow \infty$, the system reaches the disease- and fear-free equilibrium:
\begin{equation}
(S,S^F,I,R )= ( N-R_\infty,0,0,R_{\infty} ) .
\end{equation}
There is no possibility of an endemic state of  fear.
Fear can only be produced by the presence of infected people. As soon as the infection dies out, fearful people recover from their fear by interacting
with all the susceptible and recovered individuals and become susceptible themselves.

\subsection*{Model II: Global, prevalence-based spread of the  fear of the disease}
\label{model_3}
\begin{figure}
\begin{centering}
\includegraphics[width=0.5\textwidth]{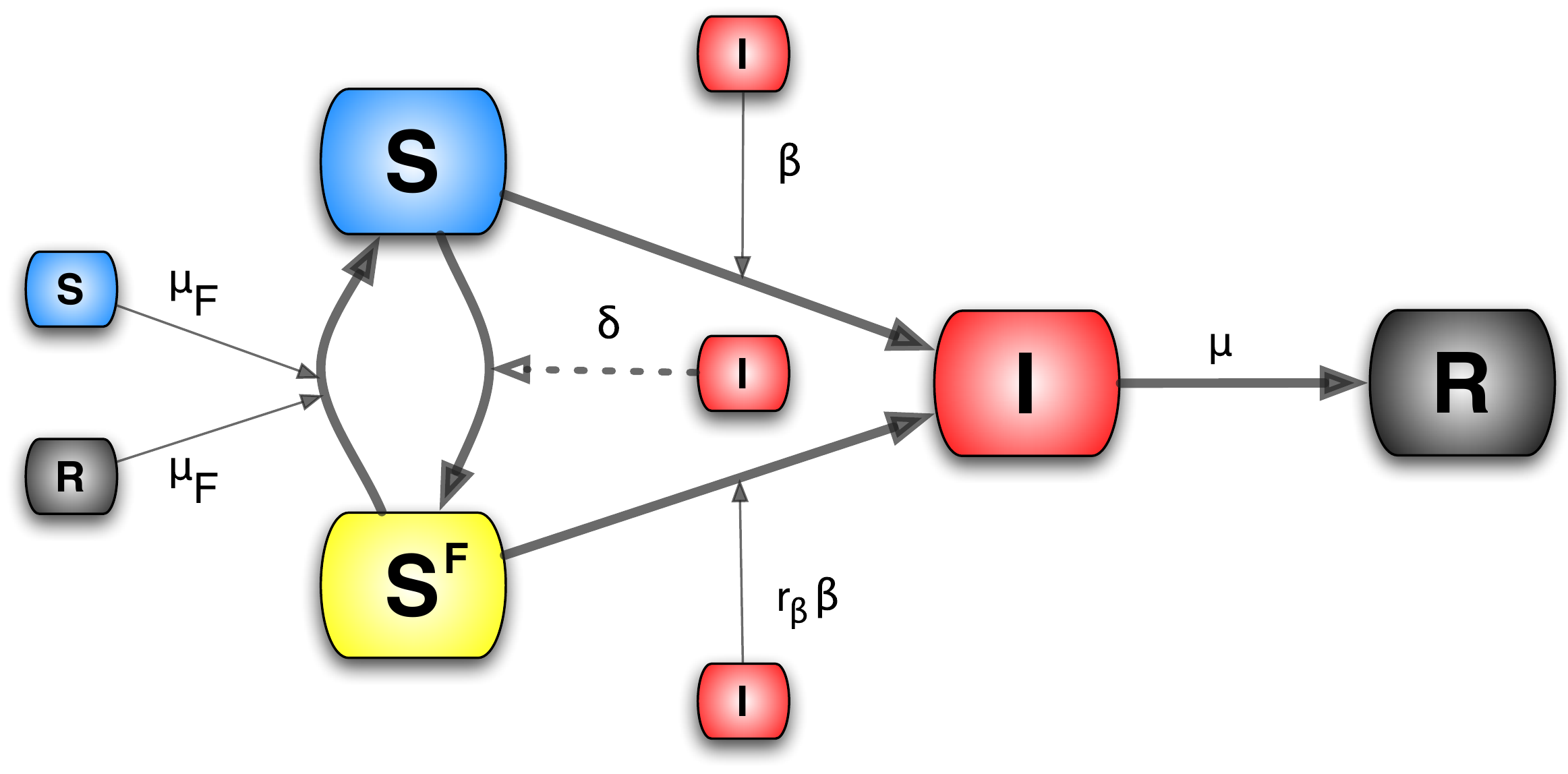}
\par\end{centering}
\caption{\label{model3} {\small {\bf Model II.} Schematic representation of Model II. The pseudo mass-action law is represented by the dashed line. }}
\end{figure}

\begin{figure}
\begin{centering}
\includegraphics[width=0.5\textwidth]{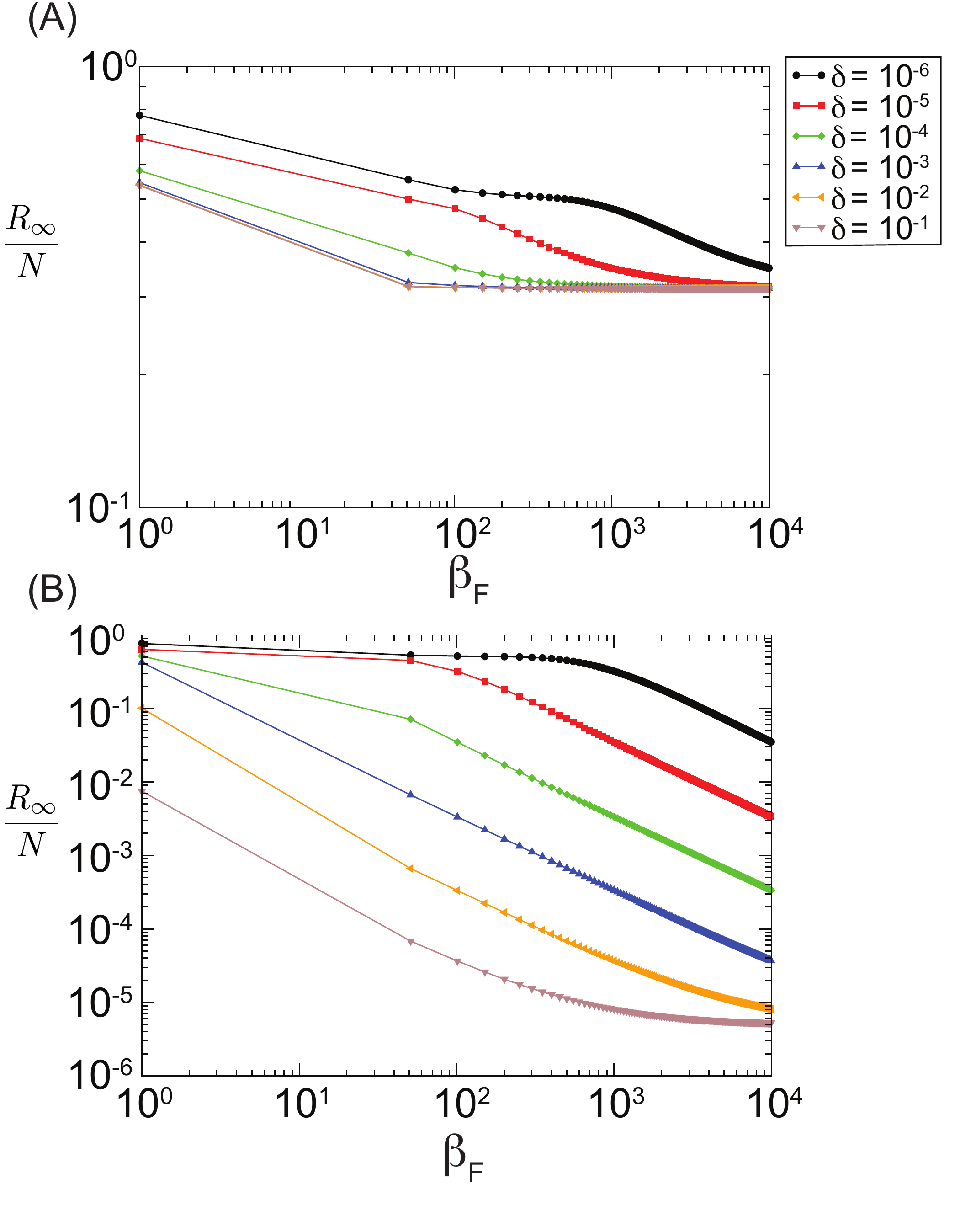}
\par\end{centering}

\caption{\label{exp_1} {\small {\bf Model II} Reduction of the epidemic size as a function
of $\beta_{F}$ for different values of $\delta$ and $r_\beta$. We fix $R_{0}=2$, $\mu=0.4\;day^{-1}$,
$\mu_{F}=0.5\;day^{-1}$ and $N=10^6$. In panel (A) we assume $r_\beta=0.6$ for which $r_\beta R_0>1$. Increasing the value of $\beta_F$ results in an asymptotic value of the epidemic size other than zero. 
In panel (B) we consider $r_\beta=0.4$. In this case, instead, $r_\beta R_0<1$. By increasing the value of $\beta_F$ the epidemic size is increasingly reduced. This effect is stronger for bigger values of $\delta$.
 The values are obtained by numerical integration of the equations. }}
\end{figure}

The second fear-inducing process we consider is the spread of the fear contagion through mass-media (Model II).
 In order to increase ratings mass-media widely advertise the progress of  epidemics,
causing even the people that have never contacted a diseased person to acquire  fear of the disease.
 In this formulation, even a very small number of infected people is enough to trigger the spread of  the fear contagion. To model this we consider a pseudo mass-action law~\cite{Keeling_Rohani}
 in which the number of infected people is not 
rescaled by the total population. Hence the absolute number of infected individuals drives  the spread. 
The transition rate peculiar to this model can be written as 
 $\lambda_{S\rightarrow S^F}^{II}=\beta_F \left[ 1- e^{-\delta I(t)}\right]$. The equations describing the system read as
\begin{widetext}
\begin{eqnarray}
d_{t}S(t) = &-&\lambda_{S \rightarrow I}S(t)- \lambda^I_{S \rightarrow S^F}S(t)-\lambda^{II}_{S \rightarrow S^F}S(t)+\lambda_{S^F \rightarrow S}S^{F}(t),\nonumber\\
d_{t}S^{F}(t) = & -&\lambda_{S^F \rightarrow I}S^{F}(t)+ \lambda^I_{S \rightarrow S^F}S(t)+ \lambda^{II}_{S \rightarrow S^F}S(t)-\lambda_{S^F \rightarrow S}S^{F}(t),\\
d_{t}I(t) & = & -\mu I(t)+\lambda_{S \rightarrow I}S(t)+ \lambda_{S^F \rightarrow I}S^F(t),\nonumber\\
d_{t}R(t) & = & \mu I(t).\nonumber
\end{eqnarray}
\end{widetext}
% \begin{eqnarray*}
% d_{t}S(t) = &-&\lambda_{S \rightarrow I}S(t)- \lambda^I_{S \rightarrow S^F}S(t) \\
% &-& \lambda^{II}_{S \rightarrow S^F}S(t)+\lambda_{S^F \rightarrow S}S^{F}(t),\\
% d_{t}S^{F}(t) = & -&\lambda_{S^F \rightarrow I}S^{F}(t)+ \lambda^I_{S \rightarrow S^F}S(t) \\
% &+& \lambda^{II}_{S \rightarrow S^F}S(t)-\lambda_{S^F \rightarrow S}S^{F}(t),\\
% d_{t}I(t) & = & -\mu I(t)+\lambda_{S \rightarrow I}S(t)+ \lambda_{S^F \rightarrow I}S^F(t),\\
% d_{t}R(t) & = & \mu I(t).
% \end{eqnarray*}
A schematic representation of the model is provided in Figure~(\ref{model3}). 
Considering Table~\ref{tab1} we can explicitly introduce all the terms,
\begin{widetext}
\begin{eqnarray}
d_{t}S(t) & = & -\beta S(t)\frac{I(t)}{N}-\beta_{F}S(t)\left[1-e^{-\delta I(t)}\right]+ \mu_{F}S^{F}(t)\left[\frac{S(t)+R(t)}{N}\right],\nonumber\\
d_{t}S^{F}(t) & = & -r_{\beta}\beta S^{F}(t)\frac{I(t)}{N}+\beta_{F}S(t)\left[1-e^{-\delta I(t)}\right]-\mu_{F}S^{F}(t)\left[\frac{S(t)+R(t)}{N}\right],\\
d_{t}I(t) & = & -\mu I(t)+\beta S(t)\frac{I(t)}{N}+r_{\beta}\beta S^{F}(t)\frac{I(t)}{N},\nonumber\\
d_{t}R(t) & = & \mu I(t),\nonumber
\end{eqnarray}
\end{widetext}
%  \begin{eqnarray*}
% d_{t}S(t) & = & -\beta S(t)\frac{I(t)}{N}-\beta_{F}S(t)\left[1-e^{-\delta I(t)}\right]+\\
% &+& \mu_{F}S^{F}(t)\left[\frac{S(t)+R(t)}{N}\right],\\
% d_{t}S^{F}(t) & = & -r_{\beta}\beta S^{F}(t)\frac{I(t)}{N}+\beta_{F}S(t)\left[1-e^{-\delta I(t)}\right]-\\
% &-& \mu_{F}S^{F}(t)\left[\frac{S(t)+R(t)}{N}\right],\\
% d_{t}I(t) & = & -\mu I(t)+\beta S(t)\frac{I(t)}{N}+r_{\beta}\beta S^{F}(t)\frac{I(t)}{N},\\
% d_{t}R(t) & = & \mu I(t).
% \end{eqnarray*}
yielding that the population size is fixed,
\begin{equation}
\sum_{i}d_{t}X_{i}(t)=0 \;\ {\rm for} \,\ \forall \,\ t \;\ {\rm and} \,\ X_{i} \in \left[ S,S^{F},I,R \right].
\end{equation}
As in the previous model, if the infection spreads faster than the fear contagion, then the reproductive number is simply
$R_0=\beta/\mu$. In the opposite limit  it is easy to understand
that the reproductive number is $R^F_{0}=r_{\beta}R_{0}$. In this latter limit, if $r_\beta R_0<1$, then the global prevalence-based spread of  fear suppresses the spread of the disease. Moreover, in general we will have 
a reduction in the epidemic size as a function of $r_\beta$. The early time progression of $S^{F}$ is analogous to that of Model I: 
\begin{equation}
S^{F}(t)\sim \frac{\delta \beta_{F}}{\mu(R_{0}-1)+\mu_{F}} \left(e^{\mu(R_{0}-1)t}-e^{-\mu_{F}t}\right).
 \end{equation}
The analogy is due to the fact that as in the first model the transition to $S^F$ is related only to the presence of infected individuals. Even in this case the condition $\mu(R_0-1)>-\mu_F$ is always satisfied 
so that if $R_0>1$, then  fear can spread in the population.\\
 Interestingly, there is a region of the phase space in which this model and Model I are equivalent. In both models the transition to fear is related only to the presence of infected individuals. In the first model 
we use a mass-action law while in the second we use a pseudo mass-action law. It is possible to relate one of the transmission rates of fear to the other by tuning the parameters.
Let us focus our attention on small values of $\delta$. We can approximate the transition rate by
\begin{equation}
\label{app_III}
\lambda_{S\rightarrow S^F}^{II}=\beta_F [\delta I(t) +\mathcal{O}(\delta^2)] .
\end{equation}
Let us consider the first order term only, i.e., $\lambda_{S\rightarrow S^F}^{II} \sim \beta_F \delta I(t)$. The relation between the two transmission rates can easily be obtained by imposing 
\begin{equation}
\lambda_{S\rightarrow S^F}^{II}=\lambda_{S\rightarrow S^F}^{I},
\end{equation}
which leads to 
\begin{equation}
\beta_F^{II}=\beta_F^{I}\frac{1}{N \delta},
\end{equation}
where we define $\beta_F^{II}$ as the rate in the second model, given $\beta_F^{I}$ in the first. The above relation guarantees the equivalence of the two models at the first order on $\delta$.
 In the   small $\delta$ region in which the approximation~(\ref{app_III}) holds, Model I and II are mathematically indistinguishable for suitable values of the parameters, which indicates that even in the phase space of Model II we have multi-peak regions. These regions, of course, coincide with
 the regions in the first model.\\
The disease-free equilibrium of this model does not allow for an endemic state of  fear,
\begin{equation}
(S_{\infty},S_{\infty}^{F},I_{\infty},R_{\infty})=(N-R_{\infty},0,0,R_{\infty}),
\end{equation}
as the transition to fear is induced by the presence
of infected individuals only. 
As soon as the epidemic dies out the in-flow to the $S^{F}$ compartment stops, while the out-flow continues to
allow people to recover from fear. When the number of infected individuals goes to zero, the media coverage vanishes, as does the fear it causes. \\
 Even in this model the effect of fear results in a reduction of the epidemic size. This reduction is a function of $\delta$ and of all of
the parameters. As $\delta$ increases the transition into fear becomes faster. Since the people in  compartment $S^F$ are more protected from the disease, the epidemic size inevitably decreases.
While keeping the value of $\delta$ fixed, increasing $\beta_{F}$ reduces the epidemic size and drives it to its asymptotic value. 
The asymptotic value
of $R_{\infty}$ as a function of $\beta_{F}$ depends on the product $r_{\beta}\beta/\mu$. If this product is bigger than $1$, obtained through direct numerical integration of the equations as shown in Figure
(\ref{exp_1})-A, then the asymptotic value is equal to the epidemic size of an SIR model with $\beta'=\beta r_{\beta}$. If  the product is  smaller
than $1$, obtained similarly through direct integration of the equation as shown in Figure~(\ref{exp_1})-B, then the asymptotic value is zero; the rate of the spread of awareness is infinitely faster than the spread of the 
disease. This dynamic can be thought as that of an SIR with a reproductive number smaller than $1$.\\

%
%\begin{figure}
%\begin{centering}
%\includegraphics[width=0.5\textwidth]{fig/two_peak_model2_2.pdf}
%\par\end{centering}
%
%\caption{\label{curves_SSF} {\small {\bf Model III:} multiple waves of infection. Fixing $\mu_{F}=0.5$,
%$r_{\beta}=0.42$, $\alpha=0.05$, $R_{0}=2$, $\mu=0.4$, $N=10^6$ we show $100$ stochastic runs and the medians evaluated considering $5 \times 10^3$ runs for two different values of $R_F$. 
%In panel (A) we have $R_F=1.2$. In panel (B) we have $R_F=1.4$.}}
%
%\end{figure}
\subsection*{Model III: Local, belief-based spread of  the fear of the disease}
\label{model_2}
\begin{figure}
\begin{centering}
\includegraphics[width=0.5\textwidth]{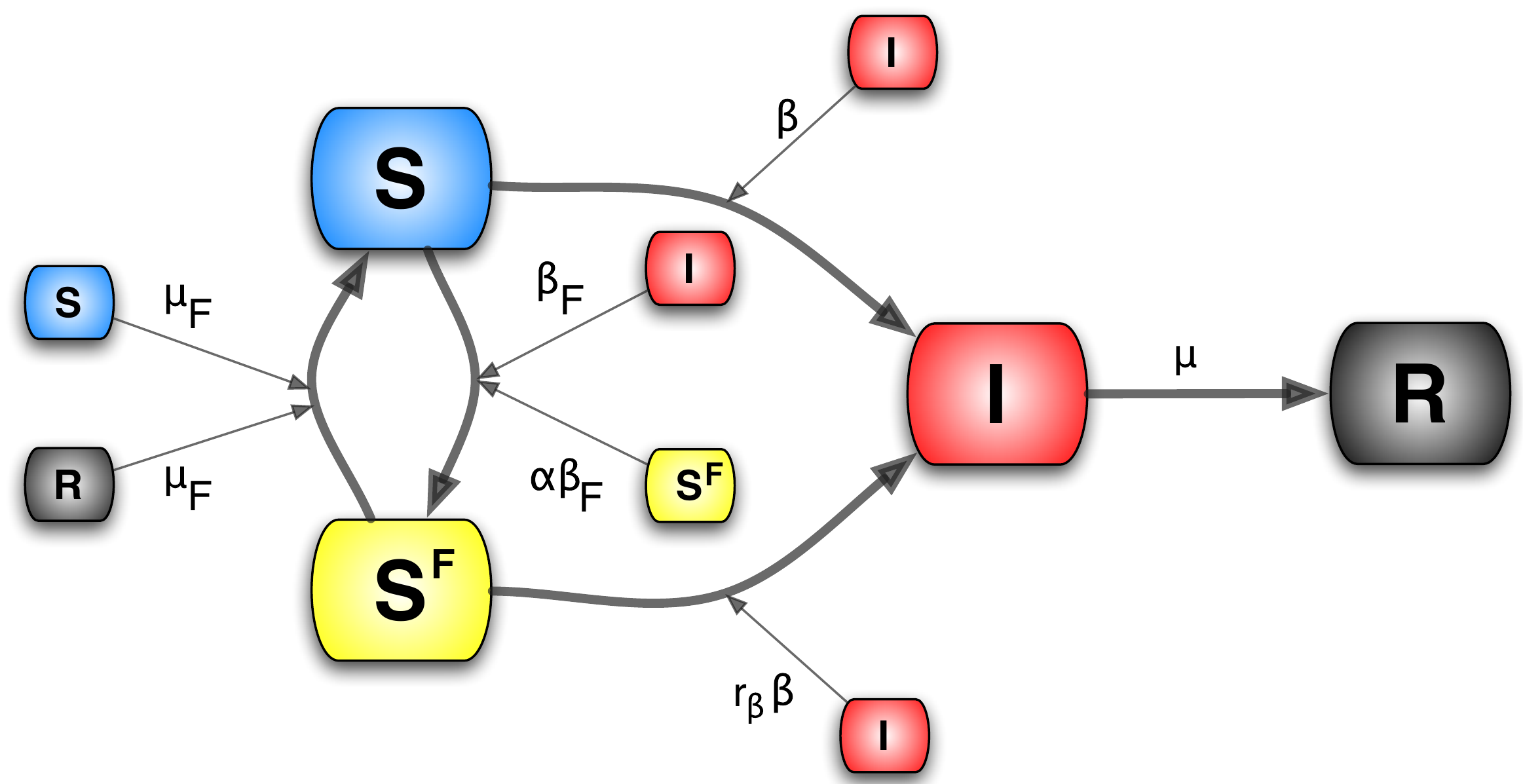}
\par\end{centering}
\caption{\label{modelII} {\small {\bf Model III.} Schematic representation of Model III. }}
\end{figure}

In this section we introduce the last model (Model III) in which we also consider self-reinforcing fear spread which accounts for 
the possibility that individuals might enter the compartment $S^F$ simply by interacting with people in this compartment:
fear generating fear. In this model people could develop fear of the infection both by interacting with infected persons and with people already concerned about the disease. A new parameter, $\alpha\geq0$,
 is necessary to distinguish between these two 
interactions. We assume that these processes, different in their nature, have different rates. To differentiate them we consider 
that people who contact infected people are more likely to be scared of the disease
than those who interact with fearful individuals. For this reason we set $0 \leq\alpha \leq1$.\\
Let us consider the case of the limit in which no infected individuals are present in the population. The $S^F$ compartment can only 
grow through the interaction $S+S^F \xrightarrow{\alpha \beta_F} 2 S^F$. It is possible to show that in the early stage this can be thought of as an SIS-like model. 
 Let us consider the case in which there are no infected individuals and just 
one individual in the compartment $S^F$, i.e., $S^F(t=0)=1$. Considering this limit, the set of equations of Model III could be written as
\begin{eqnarray*}
d_{t}S(t) & = & -\alpha\beta_{F}S(t)\frac{ S^{F}(t)}{N}+ \mu_{F}S^{F}(t)\frac{S(t)}{N},\\
d_{t}S^{F}(t) & = &  \alpha\beta_{F}S(t)\frac{ S^{F}(t)}{N}-\mu_{F}S^{F}(t)\frac{S(t)}{N},\\
d_{t}I(t) & = & 0,\\
d_{t}R(t) & = & 0.
\end{eqnarray*}
We assume that in this early stage  all the population is almost fully susceptible $S(t\sim 0) \sim N$. The equation for $S^F$ is then 
\begin{equation}
d_t S^F(t)=\alpha \beta_F S^F(t) -\mu_F S^F(t)= \left[ \alpha \frac{\beta_F}{\mu_F}-1\right ] \mu_FS^F(t).
\end{equation}
% \begin{eqnarray*}
% d_t S^F(t) &=&\alpha \beta_F S^F(t) -\mu_F S^F(t)\\
% &=& \left[ \alpha \frac{\beta_F}{\mu_F}-1\right ] \mu_FS^F(t).
% \end{eqnarray*}
This is the typical early-time term for the `infected' individuals in an SIS model. The spread of fear contagion will start if
\begin{equation}
\alpha \frac{\beta_F}{\mu_F}-1>0.
\end{equation}
This allows us to define the reproductive number of fear by
\begin{equation}
R_F \equiv \alpha \frac{\beta_F}{\mu_F}.
\end{equation}
In isolation, the fear contagion process is analogous to the reproductive number of an SIS or SIR model with transmission rate $\alpha \beta_F$. However, in the general case the spread of the fear of infection is coupled with the actual disease spread.
 The complete set of equations is
 \begin{widetext}
\begin{eqnarray}
d_{t}S(t) = &-&\lambda_{S \rightarrow I}S(t)- \lambda^I_{S \rightarrow S^F}S(t)- \lambda^{III}_{S \rightarrow S^F}S(t)+\lambda_{S^F \rightarrow S}S^{F}(t),\nonumber\\
d_{t}S^{F}(t) = & -&\lambda_{S^F \rightarrow I}S^{F}(t)+ \lambda^I_{S \rightarrow S^F}S(t)+\lambda^{III}_{S \rightarrow S^F}S(t)-\lambda_{S^F \rightarrow S}S^{F}(t),\\
d_{t}I(t) & = & -\mu I(t)+\lambda_{S \rightarrow I}S(t)+ \lambda_{S^F \rightarrow I}S^F(t),\nonumber\\
d_{t}R(t) & = & \mu I(t).\nonumber
\end{eqnarray}
\end{widetext}
% \begin{eqnarray*}
% d_{t}S(t) = &-&\lambda_{S \rightarrow I}S(t)- \lambda^I_{S \rightarrow S^F}S(t) \\
% &-& \lambda^{III}_{S \rightarrow S^F}S(t)+\lambda_{S^F \rightarrow S}S^{F}(t),\\
% d_{t}S^{F}(t) = & -&\lambda_{S^F \rightarrow I}S^{F}(t)+ \lambda^I_{S \rightarrow S^F}S(t) \\
% &+& \lambda^{III}_{S \rightarrow S^F}S(t)-\lambda_{S^F \rightarrow S}S^{F}(t),\\
% d_{t}I(t) & = & -\mu I(t)+\lambda_{S \rightarrow I}S(t)+ \lambda_{S^F \rightarrow I}S^F(t),\\
% d_{t}R(t) & = & \mu I(t).
% \end{eqnarray*}
A schematic representation of the model is provided in Figure~(\ref{modelII}). 

Considering Table~\ref{tab1} we can write all of the terms explicitly,
\begin{widetext}
\begin{eqnarray}
\label{model3-duygu}
d_{t}S(t) & = & -\beta S(t)\frac{I(t)}{N}-\beta_{F}S(t)\left[\frac{I(t)+\alpha S^{F}(t)}{N}\right]+\mu_{F}S^{F}(t)\left[\frac{S(t)+R(t)}{N}\right],\nonumber \\
d_{t}S^{F}(t) & = & -r_{\beta}\beta S^{F}(t)\frac{I(t)}{N}+ \beta_{F}S(t)\left[\frac{I(t)+\alpha S^{F}(t)}{N}\right]-\mu_{F}S^{F}(t)\left[\frac{S(t)+R(t)}{N}\right],\nonumber \\
d_{t}I(t) & = & -\mu I(t)+\beta S(t)\frac{I(t)}{N}+r_{\beta}\beta S^{F}(t)\frac{I(t)}{N}, \nonumber \\
d_{t}R(t) & = & \mu I(t).
\end{eqnarray}
\end{widetext}
% \begin{eqnarray}
% \label{model3-duygu}
% d_{t}S(t) & = & -\beta S(t)\frac{I(t)}{N}-\beta_{F}S(t)\left[\frac{I(t)+\alpha S^{F}(t)}{N}\right]+\nonumber \\
% &+& \mu_{F}S^{F}(t)\left[\frac{S(t)+R(t)}{N}\right],\nonumber \\
% d_{t}S^{F}(t) & = & -r_{\beta}\beta S^{F}(t)\frac{I(t)}{N}+ \beta_{F}S(t)\left[\frac{I(t)+\alpha S^{F}(t)}{N}\right]- \nonumber\\
% &-& \mu_{F}S^{F}(t)\left[\frac{S(t)+R(t)}{N}\right],\nonumber \\
% d_{t}I(t) & = & -\mu I(t)+\beta S(t)\frac{I(t)}{N}+r_{\beta}\beta S^{F}(t)\frac{I(t)}{N}, \nonumber \\
% d_{t}R(t) & = & \mu I(t).
% \end{eqnarray}
Also in this model we assume that the population size is fixed,
\begin{equation}
\sum_{i}d_{t}X_{i}(t)=0 \;\ {\rm for} \,\ \forall \,\ t \;\ {\rm and} \,\ X_{i} \in \left[ S,S^{F},I,R \right].
\end{equation}
If we consider the case in which the disease spreads faster than the fear of it, then
the reproductive ratio is $R_{0}=\beta/\mu$. In the opposite case  the reproductive ratio is governed by the compartment $S^F$ so that $R^F_0=r_\beta R_0$ and the epidemic size
 will be reduced depending on the value of $r_\beta$. 
In this latter case, if $r_\beta R_0<1$, then the protection from infection gained in the compartment $S^F$ causes the disease to fade out. 
Following the same linearization strategy
 shown in previous sections, the early stage of the $S^{F}$ compartment is given by
\begin{eqnarray}
S^{F}(t)&\sim&\frac{\beta_{F}}{\mu(R_{0}-1)-\mu_{F}(R_{F}-1)}\nonumber\\
&\times& \left[ e^{\mu(R_{0}-1)t}-e^{\mu_{F}(R_{F}-1)t}\right].
\end{eqnarray}
% \begin{eqnarray}
% S^{F}(t)&\sim& \frac{\beta_{F}}{\mu(R_{0}-1)-\mu_{F}(R_{F}-1)} \times \\ \nonumber
% &\times& \left[ e^{\mu(R_{0}-1)t}-e^{\mu_{F}(R_{F}-1)t}\right].
% \end{eqnarray}
Two different regions in the parameter space are then identified: one in which the rate of increase of  fear is dominated by its own thought contagion process, $\mu_F(R_{F}-1)>\mu(R_{0}-1)$, and one in which the rate of the 
local belief-based spread is dominated by the disease, $\mu(R_{0}-1)>\mu_{F}(R_{F}-1)$. 
In the first case the fear spreads independently of the value of $R_0$, and the epidemic size will be reduced due to the 
protection that individuals gain in the $S^F$ compartment.
%%%

\begin{figure}
\begin{centering}
\includegraphics[width=0.5\textwidth]{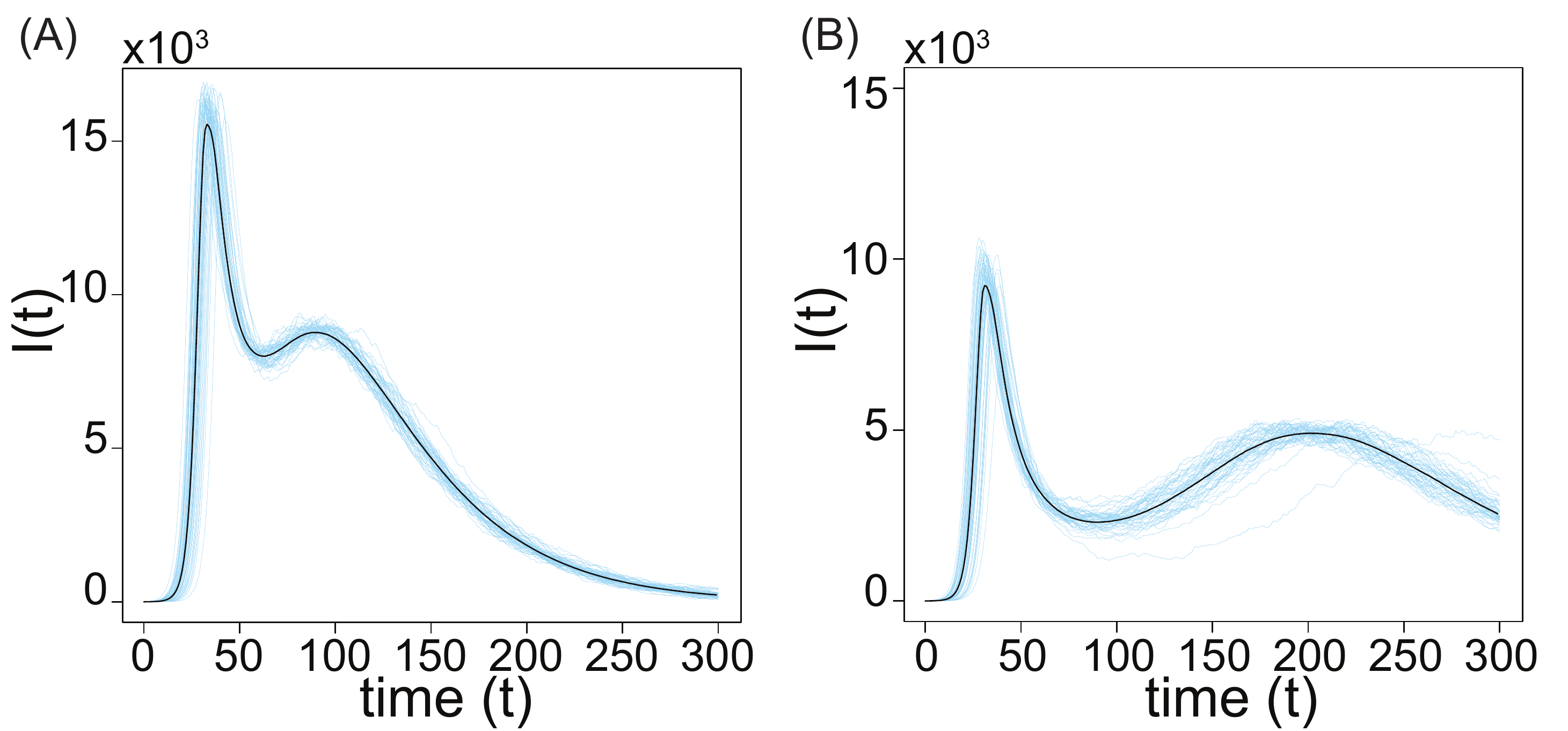}
\par\end{centering}

\caption{\label{curves_SSF} {\small {\bf Model III} Multiple waves of infection. Fixing $\mu_{F}=0.5\;day^{-1}$,
$r_{\beta}=0.42$, $\alpha=0.05$, $R_{0}=2$, $\mu=0.4\;day^{-1}$ and $N=10^6$ we show $100$ stochastic runs and the medians evaluated considering $5 \times 10^3$ runs for two different values of $R_F$. 
In panel (A)  $R_F=1.2$. In panel (B)  $R_F=1.4$.}}

\end{figure}

\begin{figure*}
\begin{centering}
\includegraphics[width=1.0\textwidth]{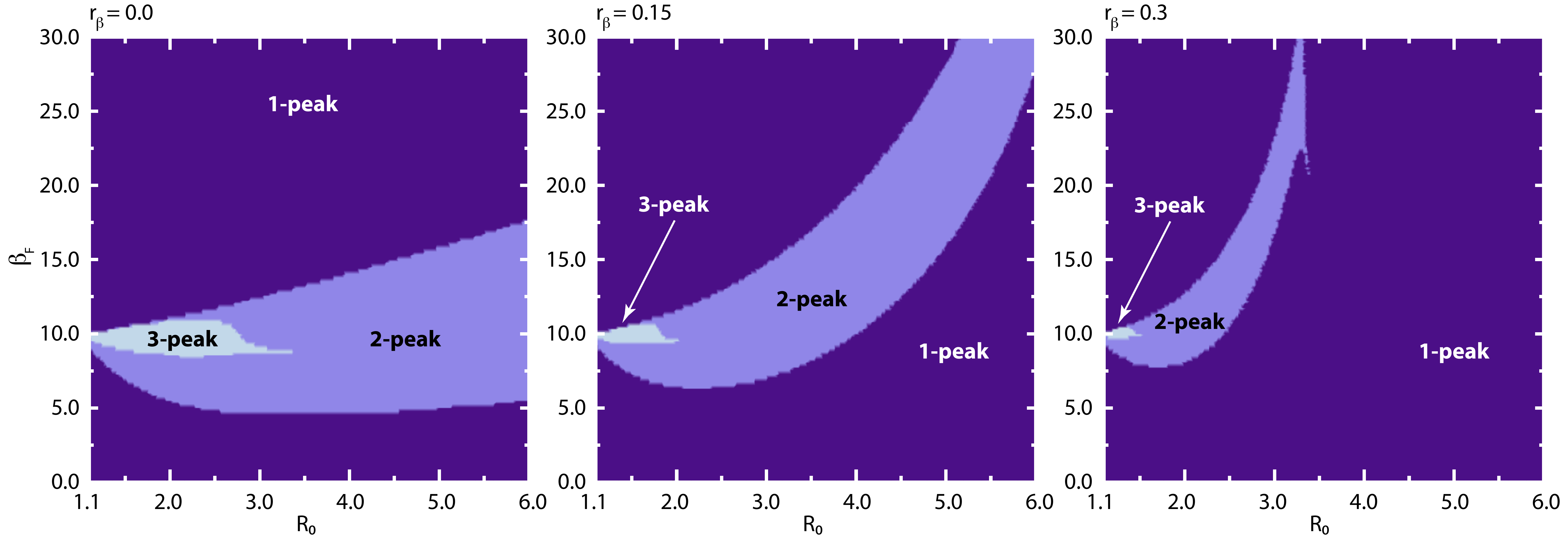}
\par\end{centering}

\caption{\label{multipeak_model3} {\small {\bf Model III} Phase diagram of infection waves on $R_0$-$\beta_F$ plane.
We display the regions of parameter space  on $R_0$-$\beta_F$ plane exhibiting different number of disease activity peaks for  three different values of $r_\beta=0, \; 0.15,\; 0.3$, where we have fixed $\mu=0.4\;day^{-1}$, $\mu_{F}=0.5\;day^{-1}$, $\alpha=0.05$ and $N=10^6$.
The phase diagram has been obtained by numerical integration of the deterministic equations in Eq.~\eqref{model3-duygu}.
}}

\end{figure*}

The new interaction, although intuitively simple, significantly complicates
the dynamics of the model. In particular within several regions of the parameter space
we observe two epidemic peaks as demonstrated in Figure~(\ref{curves_SSF}). In this figure we plot the medians for two different values of $R_F$ evaluated considering at least $5 \times 10^3$ runs in which the 
epidemic size is at least $0.1 \%$ of the population. We also show $50$ stochastic runs of the model to explicitly visualize the fluctuation among them.
This non-trivial behavior can be easily understood. Fear reinforces
itself until it severely depletes the reservoir of susceptible individuals, causing
a decline in new cases. As a result people are lured into a false
sense of security and return back to their normal behavior (recovery from
fear) causing a second epidemic peak that can  be even larger than
the first. Some authors believe that a similar process 
occurred
during the $1918$ pandemic, resulting in multiple epidemic peaks~\cite{bootsma07-1,Hatchett07-1, markel07-1}.
{
We show in Figure~(\ref{multipeak_model3}) for a set of model parameters the phase diagram of the model on $R_0$-$\beta_F$ plane regarding different number of disease activity peaks as obtained by numerical integration of the deterministic equations. The figure should be considered as illustrative as we do not have any analytical expression on the sufficient conditions yielding multiple infection peaks.  
} 
% 

% {\bf figure stays or not?} 
% A better understanding of the conditions in which this is possible
% is of obvious practical importance. In Figure~(\ref{phase_space}) we
% show the numerically found, through deterministic integration of the equations, region of parameters space that results
% in two epidemic peaks for a fixed value of $\alpha$. It is clear
% how this region is reduced and shifted toward regions of larger $R_{F}$
% and smaller $R_{0}$ as $r_{\beta}$ increases. This effect is due
% to the reduction on the protection from contagion as $r_{\beta}$ increases. In the limit
% $r_{\beta}\equiv1$ the model is indistinguishable from an SIR model
% that does not allows a second peak. 
% 
% \begin{figure}
% \begin{centering}
% \includegraphics[width=0.5\textwidth]{fig/3dplot_alpha005.pdf}
% \par\end{centering}
% 
% \caption{\label{phase_space} {\small {\bf Model II:} phase space of parameters $R_{0}\times R_{F}\times r_{\beta}$
% in which we get two peaks for $\alpha=0.05$. In the simulations we
% explored the region $1<R_{0}\leq10$, $0.1<R_{F}<20$ and $0\leq r_{\beta}<1$. The values are obtained through direct numerical integration of the equations.}}
% 
% \end{figure}
\subsubsection*{Residual collective memory of the disease and its effect on epidemic resurgence}
At the end of the disease epidemic the system enters the disease-free stage. Setting $I(t)=0$ and the epidemic size to $R_\infty$ the set of differential equations becomes
\begin{widetext}
\begin{eqnarray}
\label{disease_free}
d_{t}S(t)  = & - &\alpha \beta_F S(t)\frac{S^F(t)}{N} +\mu_{F}S^{F}(t)\left[\frac{N-S^F(t)}{N}\right],\\ \nonumber
d_{t}S^{F}(t)  = &+& \alpha \beta_F S(t)\frac{S^F(t)}{N} -\mu_{F}S^{F}(t)\left[\frac{N-S^F(t)}{N}\right],\\ \nonumber
d_{t}I(t) & = & 0,\\ \nonumber
d_{t}R(t) & = & 0.
\end{eqnarray}
\end{widetext}
% \begin{eqnarray}
% \label{disease_free}
% d_{t}S(t)  = & - &\alpha \beta_F S(t)\frac{S^F(t)}{N} +\\ \nonumber 
% &+& \mu_{F}S^{F}(t)\left[\frac{N-S^F(t)}{N}\right],\\ \nonumber
% d_{t}S^{F}(t)  = &+& \alpha \beta_F S(t)\frac{S^F(t)}{N} + \\ \nonumber 
% &-& \mu_{F}S^{F}(t)\left[\frac{N-S^F(t)}{N}\right],\\ \nonumber
% d_{t}I(t) & = & 0,\\ \nonumber
% d_{t}R(t) & = & 0.
% \end{eqnarray}
Conservation of the total number of individuals yields the following differential equation for $S^F$:
\begin{equation}
d_t S^F(t)=\mu_F\frac{S^F(t)}{N}\left[ (R_F-1)(N-S^F(t))- R_F R_\infty \right],
\end{equation}
with the  solution
\begin{equation}
S_{I=0}^{F}(t)=\frac{N\gamma}{R_{F}-1+\Theta e^{-\gamma\mu_{F}t}}.
\end{equation}
We have defined $\gamma$ as
\begin{equation} 
\gamma \equiv R_{F}\left(1- \frac{R_{\infty}}{N} \right)-1,
\end{equation}
 where $\Theta$ is a time-independent variable and is a function of the parameters of the model. Interestingly,
there are two possible disease-free equilibriums. One in which
\begin{equation}
\gamma\leq0\Rightarrow  (S_{\infty},S_{\infty}^{F},I_{\infty},R_{\infty})=(N-R_{\infty},0,0,R_{\infty}),
\end{equation}
 where fear dies along with the disease, and the one given by
\begin{eqnarray}
\label{second_d_f}
\gamma>0&\Rightarrow&(S_{\infty},S_{\infty}^{F},I_{\infty},R_{\infty})\\
&=&(\frac{R_{\infty}}{R_{F}-1},N-\frac{R_{F}R_{\infty}}{R_{F}-1},0,R_{\infty}),\nonumber
\end{eqnarray}
where fear and behavioral changes persist even after the end of the disease
epidemic. The condition $R_{F}>1$ is necessary but not sufficient in order to
have an endemic state of fear, while $R_{F}\leq1$ is sufficient to avoid an
endemic state of fear. Unfortunately, the parameter $\gamma$ is an
implicit function of the whole dynamics through the epidemic size
$R_{\infty}$.\\

The presence of an endemic state, a societal memory of the disease, and associated fear are quite  interesting features of the model induced by fear's self-reinforcement. 
In Model I  transition to the compartment $S^F$ is possible only in the presence of infected 
individuals. However, in this model  fear is able to sustain its presence in the population  if the effective reproductive number of the local belief-based spread is larger than unity even if the
disease dies out.  
Unfortunately, this argument cannot be used to fix the range of parameters in the phase space with these properties since any linearization at these stages of the compartments is not suitable. The possibility of having an endemic state of fear indicates
that an event localized in time is capable of permanently modifying
society with interesting consequences. In the case of a second epidemic,
the presence of part of the population already in the compartment
$S^{F}$ reduces the value of the basic reproduction number. 
To show this let us consider the differential equation for the infected compartment $I$ after the re-introduction of the very same infectious virus (meaning that the parameters $\beta$ and $\mu$ are equal to those of the first infectious disease):
\begin{equation}
d_t I(t)= \left [ \beta \frac{S(t)}{N} +r_\beta \beta \frac{S^F(t)}{N} - \mu \right]I(t).
\end{equation}
The initial condition of the second disease epidemic could be considered to be the disease-free equilibrium of the first epidemic. 
By using Eq.~(\ref{second_d_f}) we can express the rate equation of the infected compartment during  the early stage of the second disease as
\begin{widetext}
\begin{equation}
d_t I(t)= \left[ R_0 \frac{R_\infty}{N(R_F-1)}+  r_\beta R_0 \left( 1-\frac{R_F R_\infty}{N(R_F-1)}\right)- 1 \right] \mu I(t).
\end{equation}
\end{widetext}
% \begin{eqnarray}
% d_t I(t)&=& \left[ R_0 \frac{R_\infty}{N(R_F-1)}+ \right. \\ \nonumber
%  &+& \left. r_\beta R_0 \left( 1-\frac{R_F R_\infty}{N(R_F-1)}\right)- 1 \right] \mu I(t).
% \end{eqnarray}
Let us define $d_1\equiv R_\infty/N$ as the proportion of recovered individuals at the end of the first epidemic. In the case of the re-introduction of the  disease into the population we will have an outbreak only if the argument in the parenthesis of the above equation is larger than zero, yielding the following condition for the reproductive number $R_{0}^{II} $ of a second outbreak:
\begin{equation}
\label{red_R0}
R_{0}^{II} = \frac{\beta}{\mu}\left[r_{\beta}+\frac{d_{1}(1-r_{\beta}R_{F})}{R_{F}-1}\right] > 1.
\end{equation}
It is worth noting that the societal memory of the first outbreak increases the resistence in the population against the spread of the second outbreak in a non-trivial way.  One might be tempted to conclude that the new reproductive number is simply provided by the reproductive number of an SIR model with an equivalent proportion of removed individuals $(1-d_{1})\frac{\beta}{\mu}$, but this is not the case as we have to factor in the behavioral changes of individuals in the compartment $S^F$, obtaining 
\begin{equation}
R_{0}^{II}<(1-d_{1})\frac{\beta}{\mu}. 
\end{equation}
To prove the last inequality we have to show that
\begin{equation}
r_{\beta}+\frac{d_{1}(1-r_{\beta}R_{F})}{R_{F}-1}<1-d_{1},
\end{equation}
or
\begin{equation}
\frac{d_1(1-r_{\beta}R_{F})}{R_{F}-1}<1-d_1-r_{\beta}.
\end{equation}
The expressions on both sides of the above inequality are first-order polynomial functions of $r_{\beta}$. For $r_{\beta}=1$ they assume
the same value $-d_{1}$. It is important to stress that in this limit ($r_\beta=1$) the model is indistinguishable from the classical SIR. These two functions can only have one common point which occurs at $r_\beta=1$. We will consider only the region in which $r_\beta<1$ as assumed in our model. To prove our proposition we have to confront the slopes of the functions and show that 
\begin{equation}
\label{ine_1}
\frac{d_1R_{F}}{R_{F}-1}<1.
\end{equation}
The polynomial with smaller slope will always be below the other  in the relevant region $r_\beta<1$. Eq.~(\ref{ine_1}) can be rewritten as
\begin{equation}
d_1<1-\frac{1}{R_{F}},
\end{equation}
which is always satisfied, provided our assumption $\gamma >0$. This is an important result that confirms how an endemic state of behavioral change in the population reduces the  likelihood and impact of a second epidemic outbreak. We note that such a state will inevitably  fade out on a long time scale. This can be modeled with a spontaneous transition $S^F\to S$ acting on a time scale longer than the epidemic process itself. 

\begin{figure}
\begin{centering}
\includegraphics[width=0.4\textwidth]{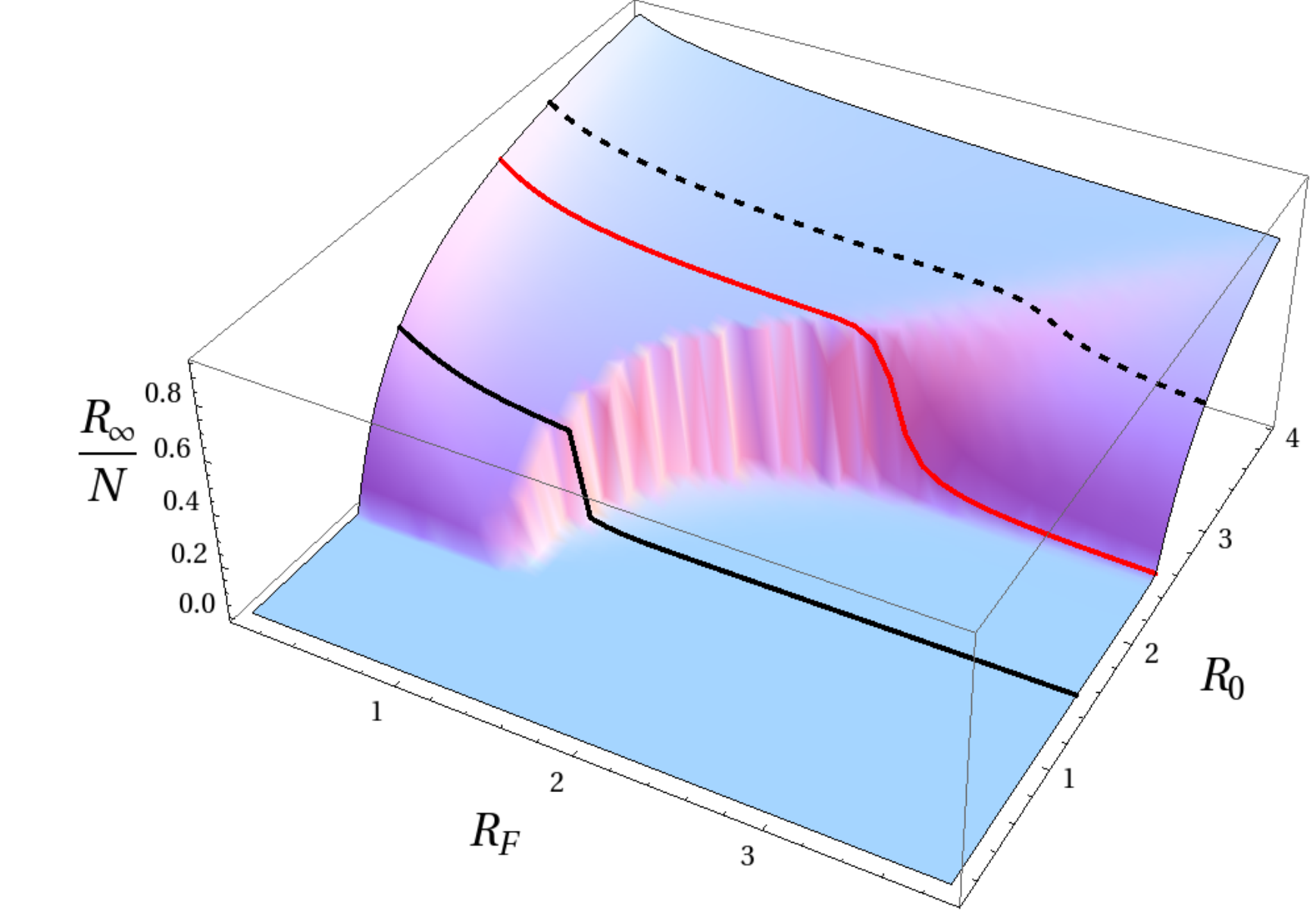}
\par\end{centering}
\caption{\label{reduction_ssf} {\small {\bf Model III} Reduction of the epidemic size as a
function of $R_{F}$ and $R_0$. 
Fixing $r_\beta=0.4$, $\mu=0.4\;day^{-1}$, $\mu_{F}=0.5\;day^{-1}$, $N=10^6$, and $\alpha=0.05$. The three lines are curves of $R_\infty/N$ as a function of $R_F$, keeping $R_0$ constant. We select three 
different values of $R_0: 1.5,2.5,3$ which correspond to solid black, red, and dashed lines, respectively. The value $R_0=2.5$ is a special case that leads to $R_0 r_\beta=1$. It divides the phase space in two different regions.
 All the values of $R_0$ below are characterized by $R_0 r_\beta <1$. In this case for large values of $R_F$ the model is reduced to an SIR with reproductive number 
$R_0 r_\beta$ below $1$ and the epidemic is halted. 
Interestingly, this behavior starts in an intermediate regime of $R_F$. There is a critical value $R_F^*$ of $R_F$ above which (i.e., $R_F> R_F^*$) the epidemic size is zero.
 This transition happens with a jump, as shown by the solid black line.  All the values of $R_0$ above $2.5$ are instead 
characterized by $R_0 r_\beta >1$. Also in this case the model is reduced to an SIR with reproductive number $R_0 r_\beta$ for large values of $R_F$,  but in this case this value is above $1$. This results in a 
epidemic size that is always non-zero. In this region of parameters no jumps are present (see the dashed line).
 The values shown in the plot are computed through numerical integration of the equations. }}
\end{figure}

\subsubsection*{Discontinuous transition in the epidemic prevalence}
\begin{figure}
\begin{centering}
\includegraphics[width=0.5\textwidth]{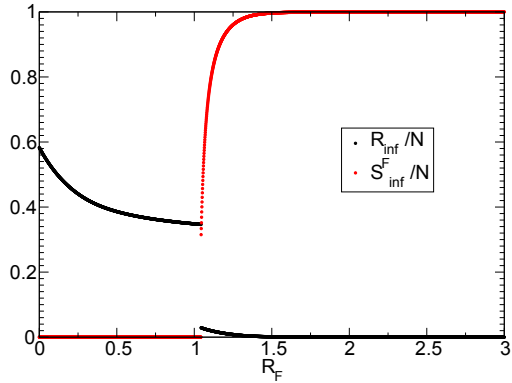}
\par\end{centering}
\caption{\label{rb_0} {\small {\bf Model III} $S_\infty^F/N$ and $R_\infty/N$ for $r_\beta=0$, $R_0=1.5$, $\mu =0.4\;day^{-1}$, $\mu_F=0.5\;day^{-1}$, $N=10^6$, and $\alpha=0.05$. The values are obtained by numerical integration of the equations. }}
\end{figure}

A further interesting characteristic of this model resides in the reduction of the epidemic size as shown in Figure~(\ref{reduction_ssf}). In this plot we show $R_\infty/N$, evaluated through direct integration of the equations,  as a function of  $R_F$ and $R_0$, keeping fixed the other parameters.
In this case the self-reinforcement  
mechanism creates a more complicated phase space that allows for a jump in the epidemic size as $R_F$ increases above a critical value $R^*_F$ (see the black solid line in Figure~(\ref{reduction_ssf})). 
This behavior, typical of the  first-order phase transitions in cooperative systems, signals a drastic change in the dynamical properties of the behavior-disease model.
If $R_F<1$, then obviously the fear of the disease is not able to affect a large fraction of the population and the disease spreads as usual in the population, affecting at the end of  its progression $R_\infty$ individuals.
 If  $R_F >1$ we face two different scenarios or two different regions of $R_0$ separated by the red solid line in Figure~(\ref{reduction_ssf}):
\begin{itemize}
\item
In the case that $R_0 r_\beta>1$ (i.e., the dashed line in Figure~(\ref{reduction_ssf})) the generation of a finite fraction of individuals in the $S^F$ compartment is not able to halt the epidemic. The behavioral changes are not enough to bring the reproductive number below the epidemic threshold and $R_\infty$ decreases smoothly because of the epidemic progress with a progressively lower effective reproductive number. 
\item
If $R_0 r_\beta\leq1$, (i.e., the black solid line in Figure~(\ref{reduction_ssf}))  the individuals that populate the $S^F$ compartment keep the spread of the epidemic  below the threshold. In principle, the state $R_\infty=0$ and $S^F=N$ would be possible. In general, the process needs to start with infectious individuals that trigger the first transitions $S\to S^F$ and therefore a small number of $R_\infty$ individuals are generated. However, there will be  a $R_F^*$ at which the growth of the fear contagion process is faster than the growth of the epidemic with a small $R_\infty$. At this point the  fear contagion process is accelerated by the growth of individuals in  $S^F$ while the epidemic spread is hampered by it.  The $S^F$ is quickly populated by individuals while the epidemic stops, generating a very small number of $R_\infty$. This generates a jump in the amount of individuals that experience the infection as a function of $R_F$.  This is clearly illustrated by Figure (\ref{rb_0}) where the behavior of both quantities $R_\infty$ and $S^F$ is plotted close to the transition point. The value at which the transition occurs also depends  on the other parameters of the model including $R_0$ and $r_\beta$. 
\end{itemize} 
The  extremely rich phase space of this model is important for two reasons: i) we have a strong reduction in the cumulative number of infected individuals associated with discontinuous transition; ii) in the case of a second epidemic the memory of the system  shifts the reproductive number towards smaller values.
 These are very interesting properties of the model due to the self-reinforcing mechanism that clearly creates non-trivial behaviors in the dynamics.
 We have tried different analytical approaches to get more insight into the phase
 transition. Unfortunately, the discontinuous transition is triggered by model behavior out of the simple linearized initial state and it is extremely difficult to derive any closed analytical expression. An analytic description is beyond the scope of the present classification of behavior-disease models and is the object of future work on the model. 

\section*{Discussion}
\label{conclu}

We introduced a general framework with different mechanisms in order to consider
 the spread of awareness of a disease as an additional contagion process. Three mechanisms were proposed. 
 In the first, basic model the social distancing effects and behavioral changes are only related to the fraction of infected individuals in the population.  
In the second we  modeled the spread of awareness considering only the absolute number of infected individuals as might happen in the case that the information the individuals rely on is mostly due to mass media reporting about the global situation. 
Finally, in the third model we added the possibility that susceptible people
 will initiate behavioral changes by interacting with individuals who have already adopted a behavioral state dominated by the fear of being infected. This apparently simple interaction allows for the self-reinforcement of fear.  
We have found that these simple models exhibit a very interesting and rich spectrum of dynamical behaviors. We have found a range of parameters with multiple peaks in the incidence curve and others in which a disease-free equilibrium is present where the population acquires a memory of the behavioral changes induced by the epidemic outbreak. 
This memory is contained in a stationary (endemic) prevalence of individuals with self-induced behavioral changes. 
Finally, a discontinuous transition in the number of infected individuals at the end of the epidemic is observed as a function of the transmissibility of  fear of the disease contagion.  At this stage the study of these properties has been mostly phenomenological and we have focused on minimal models that do not include demographic changes and spontaneous changes in the behavior of individuals such as the fading out of an epidemic over a long time. 
{
We should also note that the behavior-disease models we have suggested do not take into account the associated costs of social-distancing measures adopted by individuals, such as societal disruption and financial burden. A game theoretical approach \cite{Fenichel2011, Chen2011} would be well suited in order to account for factors in the decision making process for self-initiated behavioral changes.
}
%
%More features are being added to make the models more realistic, but, at the same time, more complex.
However, more features added to increase the realism of the models inevitably increase their complexity.
Moreover, the non-trivial dynamic behavior of the models emphasizes the importance of calibrating those features by  appropriate choices of parameter values. 
Unfortunately, in many cases we lack  the data necessary for calibrating the behavioral models. The availability of real-world, quantitative data concerning behavioral changes in populations affected by epidemic outbreaks is therefore the major roadblock to the integration of behavior-disease models. Any progress in this area certainly has  to target novel data acquisition techniques and basic experiments aimed at gathering these data. 

\section*{Acknowledgments}
This work has been partially funded by the NIH R21-DA024259 award and the DTRA-1-0910039 award to AV. The work has been also partly sponsored by the Army Research Laboratory and was accomplished under Cooperative Agreement Number W911NF-09-2-0053. The views and conclusions contained in this document are those of the authors and should not be interpreted as representing the official policies, either expressed or implied, of the Army Research Laboratory or the U.S. Government.

\end{document}